\documentclass[journal,twoside,twocolumn]{IEEEtran}
\usepackage{subfiles} 
\usepackage{cite}
\usepackage[T1]{fontenc}
\usepackage{graphicx}
\usepackage{amssymb}
\usepackage{amsmath}
\usepackage{amsthm}
\usepackage{amsfonts}
\usepackage{mathrsfs}
\usepackage{xspace}
\usepackage{bm}
\usepackage{upgreek}
\newcommand{\safemath}[2]{\newcommand{#1}{\ensuremath{#2}\xspace}}
\usepackage{setspace}
\usepackage{microtype}
\usepackage{balance}
\usepackage{xcolor}
\usepackage{framed}
\usepackage{algorithm}
\usepackage{nicefrac}
\usepackage{enumitem}
\usepackage{booktabs}
\usepackage{multirow}
\usepackage{cite}
\usepackage{adjustbox}
\usepackage{todonotes}
\usepackage{soul}
\usepackage{fancyref}
\usepackage{textcomp}

\usepackage{multirow}
\usepackage{stmaryrd}

\usepackage[caption=false,font=footnotesize]{subfig}

\allowdisplaybreaks % THIS ALLOWS EQUATIONS TO BREAK THE PAGE

\def\x{{\mathbf x}}

\def\A{{\mathbf A}}

\def\MT{{M_\text{T}}}
\def\MR{{M_\text{R}}}

\safemath{\setA}{\mathcal{A}}
\safemath{\setB}{\mathcal{B}}
\safemath{\setC}{\mathcal{C}}
\safemath{\setD}{\mathcal{D}}
\safemath{\setE}{\mathcal{E}}
\safemath{\setF}{\mathcal{F}}
\safemath{\setG}{\mathcal{G}}
\safemath{\setH}{\mathcal{H}}
\safemath{\setI}{\mathcal{I}}
\safemath{\setJ}{\mathcal{J}}
\safemath{\setK}{\mathcal{K}}
\safemath{\setL}{\mathcal{L}}
\safemath{\setM}{\mathcal{M}}
\safemath{\setN}{\mathcal{N}}
\safemath{\setO}{\mathcal{O}}
\safemath{\setP}{\mathcal{P}}
\safemath{\setQ}{\mathcal{Q}}
\safemath{\setR}{\mathcal{R}}
\safemath{\setS}{\mathcal{S}}
\safemath{\setT}{\mathcal{T}}
\safemath{\setU}{\mathcal{U}}
\safemath{\setV}{\mathcal{V}}
\safemath{\setW}{\mathcal{W}}
\safemath{\setX}{\mathcal{X}}
\safemath{\setY}{\mathcal{Y}}
\safemath{\setZ}{\mathcal{Z}}
\safemath{\emptySet}{\varnothing}
\safemath{\bA}{\mathbf{A}}
\safemath{\bB}{\mathbf{B}}
\safemath{\bC}{\mathbf{C}}
\safemath{\bD}{\mathbf{D}}
\safemath{\bE}{\mathbf{E}}
\safemath{\bF}{\mathbf{F}}
\safemath{\bG}{\mathbf{G}}
\safemath{\bH}{\mathbf{H}}
\safemath{\bI}{\mathbf{I}}
\safemath{\bJ}{\mathbf{J}}
\safemath{\bK}{\mathbf{K}}
\safemath{\bL}{\mathbf{L}}
\safemath{\bM}{\mathbf{M}}
\safemath{\bN}{\mathbf{N}}
\safemath{\bO}{\mathbf{O}}
\safemath{\bP}{\mathbf{P}}
\safemath{\bQ}{\mathbf{Q}}
\safemath{\bR}{\mathbf{R}}
\safemath{\bS}{\mathbf{S}}
\safemath{\bT}{\mathbf{T}}
\safemath{\bU}{\mathbf{U}}
\safemath{\bV}{\mathbf{V}}
\safemath{\bW}{\mathbf{W}}
\safemath{\bX}{\mathbf{X}}
\safemath{\bY}{\mathbf{Y}}
\safemath{\bZ}{\mathbf{Z}}
\safemath{\bma}{\mathbf{a}}
\safemath{\bmb}{\mathbf{b}}
\safemath{\bmc}{\mathbf{c}}
\safemath{\bmd}{\mathbf{d}}
\safemath{\bme}{\mathbf{e}}
\safemath{\bmf}{\mathbf{f}}
\safemath{\bmg}{\mathbf{g}}
\safemath{\bmh}{\mathbf{h}}
\safemath{\bmi}{\mathbf{i}}
\safemath{\bmj}{\mathbf{j}}
\safemath{\bmk}{\mathbf{k}}
\safemath{\bml}{\mathbf{l}}
\safemath{\bmm}{\mathbf{m}}
\safemath{\bmn}{\mathbf{n}}
\safemath{\bmo}{\mathbf{o}}
\safemath{\bmp}{\mathbf{p}}
\safemath{\bmq}{\mathbf{q}}
\safemath{\bmr}{\mathbf{r}}
\safemath{\bms}{\mathbf{s}}
\safemath{\bmt}{\mathbf{t}}
\safemath{\bmu}{\mathbf{u}}
\safemath{\bmv}{\mathbf{v}}
\safemath{\bmw}{\mathbf{w}}
\safemath{\bmx}{\mathbf{x}}
\safemath{\bmy}{\mathbf{y}}
\safemath{\bmz}{\mathbf{z}}
\safemath{\bmzero}{\mathbf{0}}
\safemath{\bmone}{\mathbf{1}}
\safemath{\reals}{\mathbb{R}}
\safemath{\positivereals}{\reals_{+}}
\safemath{\integers}{\mathbb{Z}}
\safemath{\posint}{\integers_{+}}
\safemath{\naturals}{\mathbb{N}}
\safemath{\posnaturals}{\naturals_{+}}
\safemath{\complexset}{\mathbb{C}}
\safemath{\rationals}{\mathbb{Q}}
\DeclareMathOperator{\tr}{tr}				% trace
\newcommand*{\fancyrefapplabelprefix}{app}		% Appendix
\newcommand*{\fancyrefthmlabelprefix}{thm}		% Theorem
\newcommand*{\fancyreflemlabelprefix}{lem}		% Lemma
\newcommand*{\fancyrefcorlabelprefix}{cor}		% Corollary
\newcommand*{\fancyrefdeflabelprefix}{def}		% Definition
\newcommand*{\fancyrefproplabelprefix}{prop}	% Proposition
\newcommand*{\fancyrefobslabelprefix}{obs}		% Observation 
\newcommand*{\fancyrefalglabelprefix}{alg}		% Algorithm
\newcommand*{\fancyrefasmlabelprefix}{asm}	    % Assumption
\newcommand*{\fancyreftbllabelprefix}{tbl}	    % Assumption

\frefformat{vario}{\fancyrefseclabelprefix}{Section~#1}
\frefformat{vario}{\fancyrefthmlabelprefix}{Theorem~#1}
\frefformat{vario}{\fancyreflemlabelprefix}{Lemma~#1}
\frefformat{vario}{\fancyrefcorlabelprefix}{Corollary~#1}
\frefformat{vario}{\fancyrefdeflabelprefix}{Definition~#1}
\frefformat{vario}{\fancyrefobslabelprefix}{Observation~#1}
\frefformat{vario}{\fancyrefasmlabelprefix}{Assumption~#1}
\frefformat{vario}{\fancyreffiglabelprefix}{Figure~#1}
\frefformat{vario}{\fancyrefapplabelprefix}{Appendix~#1} 
\frefformat{vario}{\fancyrefproplabelprefix}{Proposition~#1}
\frefformat{vario}{\fancyrefalglabelprefix}{Algorithm~#1}
\frefformat{vario}{\fancyrefeqlabelprefix}{(#1)}
\frefformat{vario}{\fancyreftbllabelprefix}{Table~#1}

\newtheorem{rem}{Remark}

\begin{document}

\title{CSI-Based Multi-Antenna and Multi-Point \\ Indoor Positioning Using Probability Fusion}

\author{Emre~G\"{o}n\"{u}lta\c{s}, 
	Eric~Lei, 
	Jack~Langerman, 
	Howard~Huang, 
	and~Christoph~Studer% <-this % stops a space
	\thanks{
		E.~G\"{o}n\"{u}lta\c{s} and E.~Lei are with the School~of Electrical and Computer Engineering, Cornell University, Ithaca, NY (e-mail: eg566@cornell.edu; el536@cornell.edu).}%
	\thanks{
		J.~Langerman  and H.~Huang are with Nokia Bell-Labs, Murray Hill, NJ (e-mail: jack.langerman@nokia-bell-labs.com; howard.huang@nokia-bell-labs.com).}
	\thanks{C. Studer is with the Department of Information Technology and Electrical Engineering at ETH Z\"{u}rich, Z\"{u}rich, Switzerland (e-mail:  studer@ethz.ch).}%
	 \thanks{The work of EG and CS was supported in part by Xilinx Inc.\ and by the US NSF under grants CCF-1652065, CNS-1717559, CNS-1955997, and ECCS-1824379. 
	 The Quadro P6000 GPU used for this research was donated by the
		NVIDIA Corporation. } 
}
\maketitle

\begin{abstract}
Channel state information (CSI)-based fingerprinting via neural networks (NNs) is a promising approach to enable accurate indoor and outdoor positioning of user equipments (UEs), even under challenging propagation conditions.
In  this  paper,  we  propose  a positioning  pipeline for wireless LAN MIMO-OFDM systems which uses uplink CSI measurements obtained from one or more unsynchronized access points (APs). For each AP receiver, novel features are first extracted from the CSI that are robust to system impairments arising in real-world transceivers. These features are the inputs to a NN that extracts a probability map indicating the likelihood of a UE being at a given grid point. The NN output is then fused across multiple APs to provide a final position estimate. We provide experimental results with real-world indoor measurements under line-of-sight (LoS) and non-LoS propagation conditions for an 80\,MHz bandwidth IEEE 802.11ac system using a two-antenna transmit UE and two AP receivers each with four antennas. Our approach is shown to achieve centimeter-level median distance error, an order of magnitude improvement over a conventional baseline.
\end{abstract}


\section{Introduction}

Positioning of mobile user equipment (UE) devices is essential for a broad range of applications, including navigation, virtual reality, asset tracking, advertising, industrial automation, and many more \cite{han2016survey,fallah2013indoorsurvey,WEN201921survey,brena2017evolutionsurvey,rashdan2020csicomparison}. 
While global navigation satellite system (GNSS) technologies enable ubiquitous positioning in outdoor environments with a view of the sky, there is no similarly ubiquitous solution for indoor environments. In addition, some of the indoor positioning applications, such as drone or robot navigation, require centimeter-level precision for executing their missions, which is not achievable using conventional GNSS. 
One class of solutions for indoor positioning uses infrastructure cameras for visible or infra-red light, viewing an object of interest which is equipped with either an active transmitter or passive reflector~\cite{armstrong2013visiblelight,kuo2014visiblelight,koyuncu2010survey,lee2014ir}. While such systems can achieve sub-centimeter accuracy, they are expensive, require unobstructed views, and may not work in rooms with bright sunlight. 

\subsection{The Challenges of Indoor Positioning using RF Signals}
An alternative to camera-based positioning methods is to use radio-frequency (RF) signals~\cite{WEN201921survey,liu2007indoorsurvey,liu2019indoorsurvey,wu2013indoor,gustaffson2005wirelesspos,sahinoglu2008ultra,studer20205g}.
However, RF-based localization techniques typically incur significant costs in calibrating the infrastructure access points (APs), depending on the required level of accuracy. 
For example, achieving meter-level accuracy with RF time-difference-of-arrival measurements requires synchronization of the APs with nanosecond accuracy and knowledge of their locations with sub-meter accuracy. 

A type of RF localization technique which has the advantage of \emph{not} requiring AP calibration is known as fingerprinting. Instead, these techniques rely on the offline creation of an empirical database that records RF measurements, such as the received signal strength indicator (RSSI) from multiple APs, as a function of the UE location. 
An algorithm then estimates the location of a UE given its RF measurements and the fingerprinting database. The database could be created with relatively low cost, for example, using an RF receiver mounted on a robot that periodically moves through a space for cleaning. 

As an alternative to RSSI measurements, fingerprinting could instead be based on estimates of channel state information (CSI), which is always required for data demodulation. For wideband MIMO-OFDM systems, complex-valued CSI can be estimated for each active subcarrier and each transmit-receive antenna pair, resulting in significantly richer measurement sets compared to RSSI. CSI-based fingerprinting has been shown to enable accurate localization in both  indoor~\cite{arnold2019novel,wang2017csi,wu2013indoor,yang2013rssi} and outdoor~\cite{ericpaper,huawei2020paper,lundpaper,larsson2015fingerprinting} applications. 
Recently, a range of CSI-based fingerprinting methods that use machine learning---rather than sophisticated geometrical models---has been proposed~\cite{zappone2019surverwireless,huawei2020paper,wang2017csi,wang2015deepfi,chen2017confi,lundpaper,arnold2019novel,wang2020bimodalcsi}.
For all of these methods, carefully designed features from CSI turn out to be critical, as real-world channels exhibit small-scale fading and wireless transceivers suffer from a number of hardware impairments. 
In addition, most of these methods rely on supervised learning to train a neural network (NN) that maps CSI to UE position, which requires dedicated measurement campaigns to acquire CSI and associated ground-truth position. 
If relative location information is sufficient, self-supervised methods known as channel charting~\cite{channelcharting,saidcc,huawei2020channelcharting} avoid expensive measurement campaigns. If CSI measurements from multiple APs are available, then the accuracy of channel charting can be improved significantly~\cite{olavmultipoint,howardcc}.  

\subsection{Contributions}
In this paper, we propose a CSI-based positioning pipeline for multiple-input multiple-output (MIMO) orthogonal frequency-division multiplexing (OFDM) wireless systems, which leverages the availability of CSI acquired at multiple MIMO links. Our positioning system can be built by using off-the-shelf hardware, e.g., using the framework put forward in~\cite{rpicsi2019}, without knowing the details of the wireless infrastructure. Our main contributions can be summarized as follows:
\begin{itemize}
	\item We propose a NN-based positioning pipeline that combines so-called \emph{probability maps} from multiple MIMO links (transmit antennas and/or APs). The proposed method does not require accurate synchronization between transmit antennas and/or APs, and minimizes the amount of location information to be transmitted to a centralized processor.
	\item We evaluate different methods to combine the probability maps, which indicate the likelihood of the UE location on a predefined grid. The proposed methods include probability conflation, Gaussian conflation, and NN-based probability fusion, which improve positioning accuracy over existing NN-based CSI fingerprinting solutions.
	\item We calculate the probability maps from CSI features using NNs. The proposed method enables probability maps with irregular grid structures. Furthermore, NN training is accomplished by assigning probabilities to multiple grid points using a minimum-variance approach.
	\item We extract CSI features that are robust to hardware impairments typically arising in IEEE 802.11ac MIMO-OFDM-based systems operating indoors. The proposed CSI features outperform existing features that were designed for massive multiuser MIMO basestations.
	\item We provide experimental results with real-world indoor channel measurements under line-of-sight (LoS) and non-LoS conditions and for multi-antenna and multi-AP measurements. Our results reveal that centimeter-level accuracy is achievable using our positioning pipeline.
	
\end{itemize}

\subsection{Relevant Prior Art}

CSI-based positioning has been studied extensively in the past; see, e.g., \cite{yang2013rssi,hu2016csi,yongsen2019survey,liu2020survey,wu2013indoor,wang2017csi,lundpaper,wang2015deepfi,chapre2014csi,arnold2019novel,chen2017confi,studer20205g} and the references therein. 
Recent work has focused mainly on neural network (NN)-based approaches~\cite{wang2017csi,lundpaper,wang2015deepfi,wang2017biloc,arnold2019novel,berruet2018csi,chen2017confi,bast2020cnncsi,li2020wireless,foliadis2021ieeecsi}, which (i) do not require geometric modeling~\cite{wang2015deepfi}, (ii) avoid storage of potentially very large CSI fingerprint databases~\cite{larryasilomar}, and (iii) have the potential to generalize to areas excluded in the training set~\cite{tenbrink2018generalizationcsi}.
In contrast to such NN-based approaches, our focus is on multi-antenna and multi-point positioning in which a NN generates what we call \emph{probability maps}, a probabilistic description of UE location measured at multiple APs and from multiple transmit antennas. These probability maps can then be fused at a centralized processor to improve positioning accuracy.

The references \cite{chen2017confi} and \cite{wang2015deepfi} propose NNs that estimate the likelihood of a UE being in a certain grid point using one-hot encoded vectors trained from a fixed number of uniformly-spaced reference positions.
NN learning is then formulated as a classification problem in which the labels correspond to grid points nearest to the UE locations.
Instead of one-hot vectors, we propose an improved strategy that enables NN training for general grid structures and for general probability mass functions (PMFs).  
To compute the PMFs, we propose an optimization-based minimum-variance approach that assigns probabilities to the individual grid points so that the expected location exactly matches the UE position. 
While the references \cite{fang2018firstprobmap,youssef2005horus,xiao2012prob} also rely on a probabilistic description of UE position, these methods perform positioning using geometrical models instead of NNs.

NN-based positioning from CSI measurements requires carefully-designed features, which are robust to small-scale fading as well as system and hardware impairments. 
The use of beamspace representations to extract the incident angles has been used in  \cite{channelcharting,saidcc,lundpaper,sun2019localicationbeamspace}. The conversion of subcarrier CSI into the delay domain to extract relative time-of-flight information has been used in \cite{lundpaper,deepak2015toa,liu2018toa,sun2019localicationbeamspace}.
The use of cross-correlation in the spatial domain and 1-dimensional autocorrelation in the delay domain has been used in \cite{channelcharting,saidcc} and  \cite{agostini2020cc,huawei2020paper,huawei2020channelcharting}, respectively. Such methods improve resilience to small-scale fading and common hardware impairments, including time synchronization errors as well as residual carrier frequency and sampling rate offsets. 
To improve robustness of our positioning pipeline to small-scale fading as well as system and hardware impairments that typically arise in IEEE 802.11ac MIMO-OFDM systems, 
we improve the CSI feature extraction pipeline in \cite{huawei2020paper} specialized for massive MU-MIMO cellular systems  by (i) taking the ``instantaneous autocorrelation'' over the antenna and delay domains (no averaging over multiple measurements),
(ii) computing the autocorrelation in the antenna domain (not just the delay domain), (iii) processing both the
real and imaginary parts (instead of taking the magnitudes), and (iv) not taking the logarithm on the resulting features.

Virtually all existing NN-based positioning methods rely on single-transmit antenna CSI measurements at a single access point (AP) or basestation (BS) possibly with multiple receive antennas. 
In contrast to these methods, we generate probability maps for each individual transmit antenna, which can then be fused to improve localization accuracy while reducing the complexity of the neural network and storage requirements. 
Multi-point localization strategies, where CSI from multiple APs or cellular BSs is combined, have been proposed in \cite{olavmultipoint,howardcc} for channel charting.
While such approaches only enable relative localization, they also require the exchange of CSI features to a centralized processor that performs channel charting. 
In contrast to these approaches, we propose new methods that fuse probability maps generated at multiple APs (and from multiple transmit antennas), which not only improves (absolute) positioning accuracy but also reduces the amount of information that must be transferred to a centralized processor. 

Probability fusion of multiple sensor data such as camera images \cite{acharya2011fusion} is a widely studied subject; see, e.g., \cite{1990sensorfusion,aeberhard2011fusion,wen2017ensemble}. 
Existing outdoor positioning systems, such as GNSS, triangulation-based methods in cellular systems \cite{larsson2017triang}, or indoor positioning systems such as WorldViz \cite{worldviz} or VICON \cite{vicon}, already fuse at least three different data sources to produce a robust position estimate.
The methods proposed in this paper combine probabilistic sensor fusion with CSI-based positioning. Concretely, our methods fuse multiple probability maps that indicate the likelihood of a UE being at a given grid point using theoretically principled conflation methods put forward in \cite{hill2008conflations}. To the best of our knowledge, no work has described the use of such conflation methods in a wireless positioning application.

\subsection{Notation}

Lowercase boldface letters, uppercase boldface letters, and uppercase calligraphic letters denote column vectors, matrices, and sets, respectively.
For a matrix~$\bA$, we denote its transpose by~$\bA^T$, its Hermitian transpose by~$\A^H$, its $i$th row and $j$th column by $A_{i,j}$, and its $i$th column by~$\bm{a}_i$.
For a vector~$\bma$, the $k$th entry is denoted by~$a_k$, the $\ell^2$ and $\ell^1$ norms are $\|\bma\|_2=\sqrt{\sum_{k}|a_k|^2}$ and $\|\bma\|_1=\sum_{k}|a_k|$, respectively, and
the real and imaginary parts are $\Re(\bma)$ and $\Im(\bma)$, respectively. 
\subsection{Outline}
The remainder of this paper is organized as follows.
\fref{sec:section2} describes the operation principle of the proposed positioning pipeline and introduces the MIMO-OFDM channel model. 
\fref{sec:fingerprinting} details NN-based positioning along with CSI-feature construction. 
\fref{sec:probabilityfusion} proposes different probability fusion methods for multi-antenna and multi-point data. 
\fref{sec:results} shows results for real-world indoor channel measurements. 
\fref{sec:conclusions} concludes the paper.

\section{Operation Principle and System Model}
\label{sec:section2}

Our objective is to estimate the position of a UE from CSI measurements acquired at one or multiple APs.
We first outline the operation principle of the positioning pipeline and then  describe the multi-point MIMO-OFDM system model. 

\subsection{Operation Principle}

\begin{figure}[tp]
	\begin{center}
		\includegraphics[width=\columnwidth]{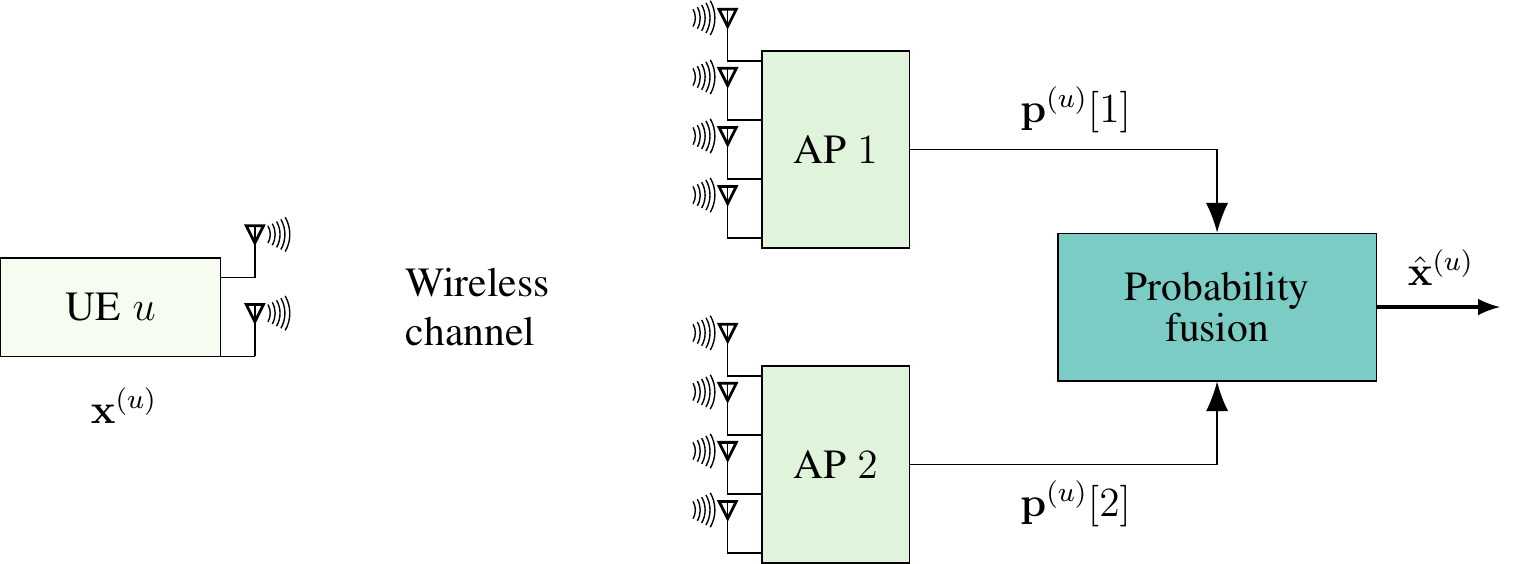}
	\end{center}
	\caption{Overview of the proposed CSI-based multi-antenna and multi-AP positioning system. A user equipment (UE) $u$ at position~$\bmx^{(u)}$ transmits pilots to one or multiple APs, which extract channel state information (CSI). Each AP then uses CSI to generate a probability map indicating the UE's location, which is fused to compute an estimate $\hat\bmx^{(u)}$ of the UE's position.}
	\label{fig:systemmodel} 
\end{figure}

\fref{fig:systemmodel} illustrates the basic concept of the proposed multi-antenna multi-point CSI-based positioning pipeline. 
The $u$th UE at location~$\bmx^{(u)}$ transmits pilot sequences from one or multiple transmit antennas, which are then used to estimate the wireless channel at one or multiple APs indexed by $b=1,\ldots,B$ where $B$ is the number of APs.

The APs can have one or many receive antennas. The acquired CSI is then used to generate a probability map $\bmp^{(u)}[b]$ which contains information on the location  of the $u$th UE as seen from the $b$th AP. 
The probability map is generated from measured CSI using neural networks, which have been trained from a dataset containing CSI-location pairs in a dedicated training phase. Details on the positioning pipeline are provided in \fref{sec:fingerprinting}. 
Ground-truth location information for neural network training is acquired via a state-of-the-art multi-camera positioning system. Details on the measurement setup are provided in \fref{sec:measurementsetup}.
The probability information from all APs is then fused in order to produce an estimate~$\hat\bmx^{(u)}$ of the UE location.

\begin{rem}
	A key limitation of supervised positioning methods is the requirement of a dedicated training phase used to acquire a large CSI dataset annotated with ground-truth position. 
	If the cost of creating a CSI dataset is manageable, then it is highly beneficial to use supervised CSI-based fingerprinting methods. 
	Self- or semi-supervised methods that build on channel charting~\cite{channelcharting,ericpaper,penghzipaper}, can be used to reduce or completely avoid the acquisition of ground-truth location information at the expense of positioning accuracy. An extension of the proposed positioning pipeline to such self- or semi-supervised methods is part of ongoing work. 
\end{rem}

\begin{rem}
	In order to acquire CSI at multiple APs, CSI sniffing methods that receive all transmitted OFDM packets in the receiver's range can be used~\cite{rpicsi2019}.
	Furthermore, the APs performing localization do not need to be the destination of the payload as CSI can be extracted from any IEEE 802.11ac~\cite{ieee80211ac}-compatible receiver. 
\end{rem}

\subsection{Multi-Antenna Multi-Point OFDM System Model}

We focus on an IEEE 802.11ac-based wireless communication system \cite{ieee80211ac}, which is also what we used in \fref{sec:results} to generate our indoor positioning results.\footnote{Many of the proposed techniques can be adapted easily to other MIMO-OFDM-based wireless communication systems.}
We consider a multi-antenna UE with $\MT$ transmit antennas that is located in the range of $B$ multi-antenna APs (or basestations) having~$\MR$ antennas each. 
We consider an OFDM system with $W$ subcarriers and  cyclic prefix length $C$. The set $\Omega_\text{used}$ contains the indices of subcarriers associated with tones that have been trained (corresponding to data and pilot subcarriers); the set~$\Omega_\text{zero}$ contains the indices associated with unused subcarriers. Consequently, we have $\Omega_\text{used} \cup \Omega_\text{zero} = \{1,\ldots,W\}$.
We assume that the $u$th UE is at position $\bmx^{(u)}\in\reals^D$, where~$D$ is typically two or three (representing the spatial location), and is transmitting one pilot per used subcarrier 
(e.g., during the preamble), which is performed at each AP to compute an estimate of the wireless channel in the frequency domain.
The estimated $\MR\times\MT$ MIMO channel matrix at subcarrier $w=1,\ldots,W$ and at AP $b=1,\ldots,B$ is denoted by $\bH^{(u)}_w[b]\in\complexset^{\MR\times\MT}$. We call the collection of these channel matrices 
\begin{align} \label{eq:CSI}
	\bH^{(u)}_w[b],\quad w=1,\ldots,W, \quad b=1,\ldots,B,
\end{align}
the CSI associated with UE $u$; we abbreviate \fref{eq:CSI} by $\{\bH^{(u)}_w[b]\}$. 
In what follows, we assume that the delay spread of the channel plus the maximum timing offset does not exceed the cyclic prefix length. We furthermore assume $C\leq|\Omega_\text{used}|$ for reasons discussed in \fref{sec:beamspace_delaydomain_transform}. Besides that, we allow the extracted channel estimates to be affected by real-world system and hardware impairments, such as timing errors and residual phase offsets.

\begin{rem}
	We do not require (i)  the $B$ APs to acquire the CSI from the $u$th UE at the same time instant or  (ii) perfect synchronization among APs. The only assumption is that the CSI measured at each APs is for the same UE, which is transmitting from approximately the same location $\bmx^{(u)}$; this includes the cases of (i) unsynchronized APs or (ii) situations in which the UE is transmitting to multiple APs in a round-robin fashion or scenarios in which one or multiple APs are acquiring CSI without decoding the UEs data. 
\end{rem}
\section{CSI-Based Fingerprinting via Neural Networks}
\label{sec:fingerprinting}

\begin{figure}[tp]
	\centering
	\subfloat[]
	{
		\adjustbox{max width=0.24\columnwidth,valign=t}{%
			\includegraphics[width=\columnwidth]{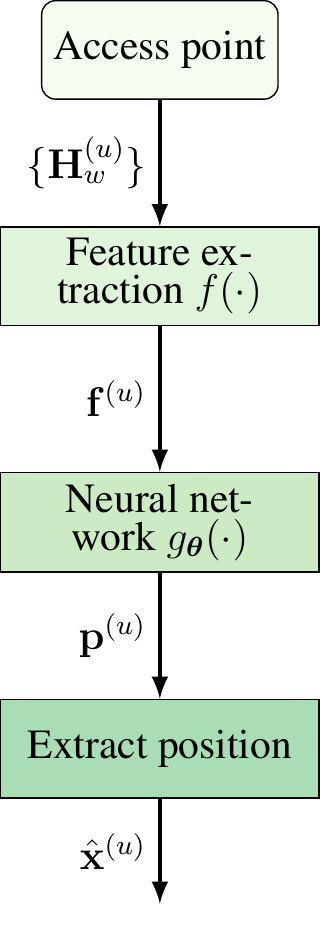}
		}
	}
	~~~~~
	\subfloat[]
	{
		\adjustbox{max width=0.557\columnwidth,valign=t}{%
			\includegraphics[width=\columnwidth]{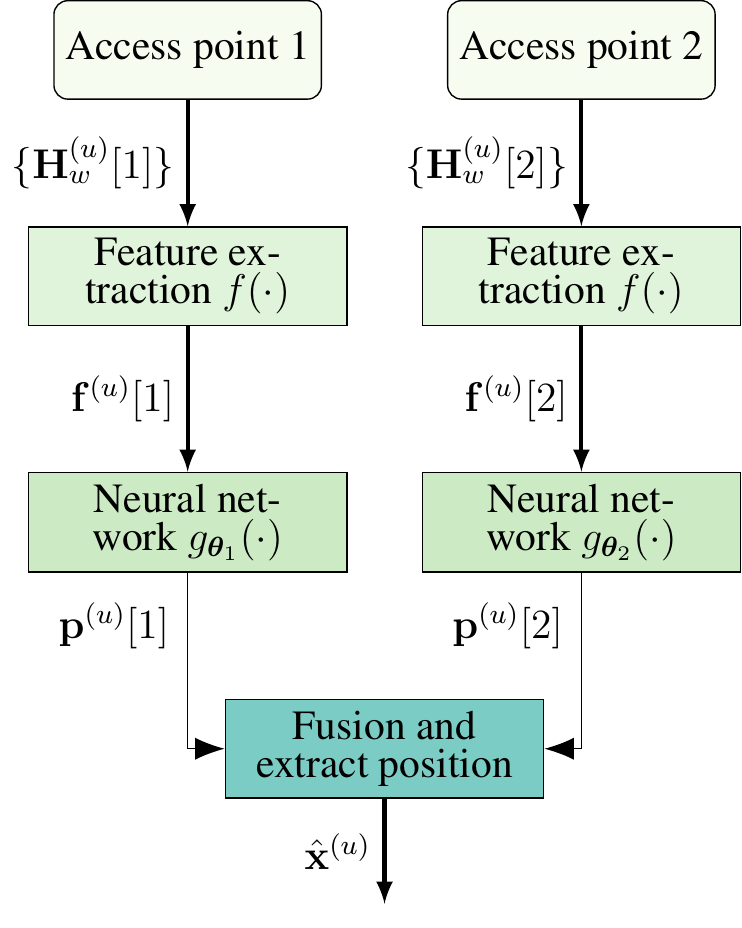}
		}
	}
	\caption{CSI-based positioning pipeline. (a) Positioning with one AP: The AP extracts CSI $\{\bH^{(u)}_w\}$ of the $u$th UE followed by a feature extraction stage that produces $\bmf^{(u)}$. A neural network then computes a probability map~$\bmp^{(u)}$ that is used to generate  an estimate $\hat\bmx^{(u)}$ of the UEs location. (b) Positioning with two APs: A centralized node fuses the two extracted probability maps $\bmp^{(u)}[1]$ and $\bmp^{(u)}[2]$ computed from two different neural networks from AP1 and AP2, respectively, to generate an estimate $\hat\bmx^{(u)}$ of the UE location.}
	\label{fig:positioningpipeline}
\end{figure}

We now detail the proposed multi-antenna multi-point CSI-based positioning pipeline illustrated in \fref{fig:positioningpipeline}.
We start by describing the CSI-feature extraction stage followed by discussing the NN that generates probability maps. Means to fuse probability maps to obtain accurate location estimates from multi-antenna and multi-point data are detailed in \fref{sec:probabilityfusion}.

\subsection{CSI-Feature Extraction}
\label{sec:CSIfeatures}
In order to enable CSI-based positioning, it is critical to construct \emph{robust} CSI-features that (i) are unique for a given UE location, (ii) are robust to small-scale fading effects \cite{channelcharting,larsson2015fingerprinting,saidcc,ericpaper,penghzipaper}, and (iii) are resilient to real-world system and hardware impairments \cite{huawei2020channelcharting,huawei2020paper}. 
Clearly, for a given UE location $\bmx^{(u)}$, the CSI features extracted from the estimated CSI  $\{\bH^{(u)}_w[b]\}$ should be unique---otherwise, UE location is not uniquely determined. 
Furthermore, small-scale fading, e.g., caused by moving objects in the surrounding area of the UE and APs, should not affect the CSI features as they are generally difficult to model. 
Finally, system and hardware impairments, such as varying transmit or receive power as well as timing, carrier frequency, and sampling rate offsets, should not affect the CSI features as they may vary over time and from UE to UE.
We emphasize that resilience to system and hardware impairments is crucial in order to achieve accurate positioning from CSI features.
We now propose such CSI features (see \fref{fig:positioningpipeline}), which address the desired properties (ii) and (iii), as CSI-feature uniqueness in property (i) is mainly determined by the physical channel and difficult to control in practice.

\subsubsection{Delay-Domain Transform}
\label{sec:beamspace_delaydomain_transform}
We start by transforming the frequency (or subcarrier) domain into the delay domain, as OFDM-based systems typically have only a limited delay spread (e.g., no larger than the cyclic prefix length)  and only a few taps in the impulse responses are significant. %
Let us first define the vector $\bmh_{m,w}^{(u)}[n] = \big[\bH^{(u)}_w[b]\big]_m$ as the $m$th column of $\bH^{(u)}_w[b]$, which is the CSI vector corresponding to the $m$th transmit antenna of the $u$th UE at subcarrier~$w$.
Next, we define the following $M$-dimensional frequency-domain vector 
\begin{align} \label{eq:frequencydomainvector}
	\hat\bmh_{n,m}^{(u)}[b] = \big[\,    [\bmh^{(u)}_{m,1}[b]]_n, [\bmh^{(u)}_{m,2}[b]]_n , \ldots, [\bmh^{(u)}_{m,W}[b]]_n  \,\big]^T
\end{align}
for a given receive antenna $n$ of AP $b$ and a given UE $u$ transmit antenna~$m$. In words, the vector $\hat\bmh_{n,m}[b]$ contains the channel estimates over all $W$ subcarriers. 
Ideally, by taking the inverse DFT of the frequency-domain vector $\hat\bmh^{(u)}_{n,m}[b]$, one would obtain the delay-domain description of the frequency-selective channel from the $m$th UE antenna to the $n$th receive antenna of AP $b$. Unfortunately, only the subcarriers indexed by $\Omega_\text{user}$ are available in practice, whereas the entries pertaining to the subcarriers indexed by~$\Omega_\text{zero}$ are generally unknown (as they were not trained). 
Fortunately, since we assumed that the cyclic prefix length $C$ is not larger than the number of used subcarriers $|\Omega_\text{used}|$, one can estimate the delay-domain coefficients within the cyclic prefix length. Let $\Gamma_\text{cp}$ be the set of indices associated to the channel taps in the delay-domain so that $C=|\Gamma_\text{cp}|$. Then, the delay-domain coefficients can be estimated by computing 
\begin{align} \label{eq:delay_estimation}
	\bmt^{(u)}_{n,m}[b] = (\bD_{\Omega_\text{used},\Gamma_\text{cp}})^\dagger \big[\hat\bmh_{n,m}^{(u)}[b]\big]_{\Omega_\text{used}}.
\end{align}
Here, the delay-domain vector $\bmt^{(u)}_{n,m}[b]  \in\complexset^{|\Gamma_\text{cp}|}$ contains the~$C$ dominant taps of the wireless channel between the $m$th transmit antenna of UE $u$ and the $n$th receive antenna at AP $b$, the matrix $\bD_{\Omega_\text{used},\Gamma_\text{cp}}$ contains the rows indexed by $\Omega_\text{used}$ and the columns indexed by $\Gamma_\text{cp}$ of the DFT matrix, $(\cdot)^\dagger$ denotes the left-pseudoinverse, and $[\hat\bmh_{n,m}^{(u)}[b]]_{\Omega_\text{used}}$ is the subset of the frequency-domain vector in~\fref{eq:frequencydomainvector} corresponding to the used subcarriers indexed by $\Omega_\text{used}$. 
Note that the least-squares estimator in \fref{eq:delay_estimation} is frequently used to denoise channel vectors in OFDM systems \cite{haene2007ofdm}.

\begin{rem}
	We have observed that taking the inverse DFT directly over the frequency-domain vector \fref{eq:frequencydomainvector} still works well in practice and requires lower complexity than the approach in \fref{eq:delay_estimation} as one can simply use an inverse fast Fourier transform (IFFT). 
\end{rem}

\subsubsection{Autocorrelation}
\label{sec:autocorrelation}
In addition to transforming the estimated CSI into the delay domain,
we now propose a method that renders the CSI features robust to time-synchronization errors, residual carrier frequency offset (CFO) and sampling rate offset (SRO), and global phase modulations.
The method proposed here is inspired by the approach proposed recently in \cite{huawei2020paper} for CSI-based positioning in cellular massive MIMO systems. 

Time synchronization errors and residual phase errors can be modeled in the discrete-time domain as $t[k]=y[k-\delta]e^{j\varphi}$, where $t[k]$ is the time-domain signal at sample index $k$, $y[k-\delta]$ is the true received signal with unknown delay~$\delta$ caused by synchronization (or frame-start detection) errors, and $\varphi$ determines the amount of residual phase errors.\footnote{Practical receivers estimate and compensate CFO and SRO, but residual phase errors remain in practice \cite{schenk2008rf,studer2010rfimpairment}.}
When computing the ``instantaneous\footnote{We are not taking any expectation over the product $t[k]t^*[k+\tau-1]$, which is in contrast to the method proposed in \cite{huawei2020paper}.} autocorrelation'' of the signal $t[k]$, we have 
\begin{align} \label{eq:autocorrelation}
	R_t[\tau]  = \! \sum_{k} t[k]t^*[k+\tau-1] 
	= \! \sum_{k'} y[k'] y^*[k'+\tau-1],
\end{align}
for $\tau=1,2,\ldots$, which does no longer depend on the time synchronization error $\delta$ and the residual phase errors.

To improve robustness of our CSI features, we will follow this approach and  compute the instantaneous autocorrelation not only in the delay domain but also in the antenna domain. The latter renders the proposed CSI-features robust to global phase offsets across the antenna array.
Let $\big[\bmt^{(u)}_{n,m}[b]\big]_k$ be the $k$th delay-domain sample measured at the $n$th receive antenna of AP~$b$ transmitted from the $m$th antenna from UE~$u$. Then, we compute  
\begin{align}
	& R^{(u)}_t[m,\tau,\kappa,b]  = \sum_{n=1}^\MR \sum_{k=1}^{C} \big[\bmt^{(u)}_{n,m}[b]\big]_k \big[\bmt^{(u)}_{n+\kappa-1,m}[b]\big]^*_{k+\tau-1},  \label{eq:autocorrelation_features}
\end{align}
where $\tau=1,2,\ldots,2C$ and $\kappa=1,2,\ldots,2\MR$.
For a given AP $b$ and UE $u$, we vectorize the three-dimensional tensor in \fref{eq:autocorrelation_features} so that the vector $\bmr^{(u)}[b]\in\complexset^{\MR\MT2C}$ contains all the entries of $R^{(u)}_t[m,\tau,\kappa,b]$ for $m=1,\ldots,\MT$, $\tau=1,\ldots,2C$, and $\kappa=1,\ldots,2\MR$.
Finally, to enable the use of off-the-shelf deep-learning software, we convert the complex valued vector $\bmr^{(u)}[b]\in\complexset^{\MT2C2\MR}$ into a $2\MT2C2\MR$-dimensional real-valued CSI-feature vector as follows:
\begin{align} \label{eq:CSIfeatures}
	\hat\bmf^{(u)}[b] = \Big[  \Re\{\bmr^{(u)}[b]\}^T , \Im\{\bmr^{(u)}[b]\}^T \Big]^T.
\end{align}

\subsubsection{CSI Feature Normalization}
In practice, the gains of the power amplifier at the UE side and the low-noise amplifier at the AP side can be set independently. While these gain settings are typically kept constant during transmission of one OFDM frame, the AP does, in general, not know the UE's gain settings. Furthermore, the path-loss characteristics in indoor applications depend on the environment. Hence, the receive power is not a reliable indicator for the distance between the UE and the AP, and should be ignored. For these reasons, we normalize the vector in \eqref{eq:CSIfeatures} as $ \bmf^{(u)}[b]  = {\hat \bmf^{(u)}[b] }/{\|\hat \bmf^{(u)}[b] \|_2}$, which ensures that the transmit and receive gains as well as the path loss are ignored, and also improves the convergence of stochastic gradient descent.

In what follows, we will also use CSI features extracted separately per AP $b$ and per transmit antenna $m=1,\ldots,\MT$, which are obtained by vectorizing $R^{(u)}_t[m,\tau,\kappa,b]$ in only $\tau=1,\ldots,2C$ and $\kappa=1,\ldots, 2\MR$ and by stacking the real and imaginary parts in the vector $\bmf^{(u)}_m[b]$. 

\begin{figure}[tp]
	\centering
	\adjustbox{max width=\columnwidth,valign=t}{%
		\includegraphics[width=\columnwidth]{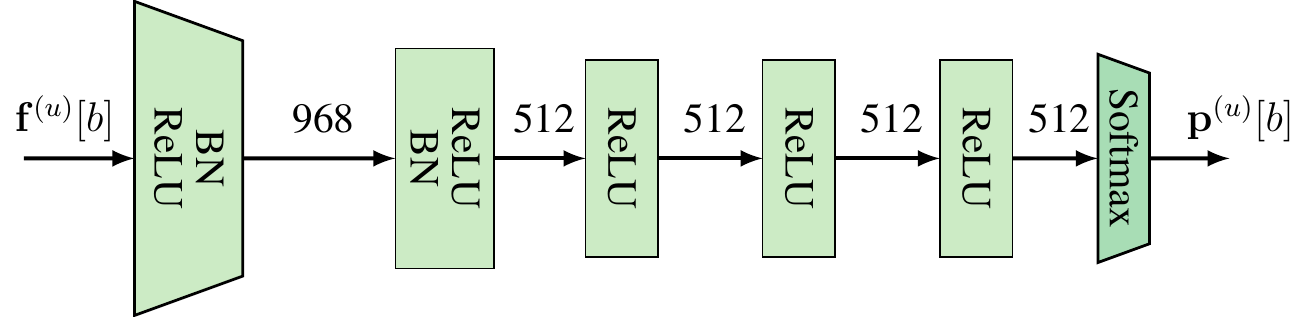}
	}
	\caption{Neural network (NN) structure at AP $b$. The NN $g_{\bm{\theta}_b}$ with weights and biases contained in the vector $\bm{\theta}_b$ takes in a CSI-feature vector $\bmf^{(u)}[b]$ and generates a probability map $\bmp^{(u)}[b]$ that describes the position of UE $u$.}
	\label{fig:nnmodel} 
\end{figure}

\subsection{Neural Network Structure}
\label{sec:nnstructure}
As illustrated in \fref{fig:positioningpipeline}, we propose to use one or multiple neural networks (NN) $g_{\bm{\theta}_b}$ at each AP $b$ with weights and biases from all layers contained in the vector $\bm{\theta}_b$ that takes in CSI  feature vectors $\bmf^{(u)}[b]$ and generates what we call a \emph{probability map}
\begin{align}
	\bmp^{(u)}[b] = g_{\bm{\theta}_b}(\bmf^{(u)}[b])
\end{align}
that describes the location of UE $u$; see \fref{sec:probabilitymaps} for details on probability maps. This probability map is then fused with probability maps from other APs to extract an estimated position $\hat\bmx^{(u)}$ of UE $u$.
The neural network structure is illustrated in \fref{fig:nnmodel}. We consider a six-layer neural network with input CSI-feature vector $\bmf^{(u)}[b]$ and output probability-map vector $\bmp^{(u)}[b] $. All but the last layers use ReLu activations; the last layer uses a softmax activation for the reasons detailed below. The first and second layer use batch normalization (BN).
The number of activations for each layer is shown in \fref{fig:nnmodel}; the dimensions of the input CSI-feature vector $\bmf^{(u)}[b]$ and the output probability-map vector $\bmp^{(u)}[b]$ depend on the feature type and the resolution of the probability map, respectively. 
Prior to training, We initialize the weights and biases using Glorot \cite{glorot} with a uniform distribution; during training,  we use the binary cross entropy (BCE) loss.

\subsection{Probability Maps}
\label{sec:probabilitymaps}
Instead of using NNs that directly produce an estimate~$\hat\bmx^{(u)}$ for the true location $\bmx^{(u)}$ of UE~$u$, which is the de-facto standard approach \cite{zappone2019surverwireless,huawei2020paper,wang2017csi,wang2015deepfi,lundpaper,arnold2019novel}, we propose to use a probabilistic description of UE location. This strategy has the advantage that the NN output contains valuable information for during downstream processing, e.g., (i) to extract reliability information on the estimated UE location and (ii) to fuse multiple probability maps to improve positioning accuracy in multi-antenna and/or multi-AP scenarios.
In addition, a probabilistic location description can also resolve ``conflicts,'' which may arise if two spatial locations generate similar  CSI. Finally,  probability maps can also improve NN training for CSI-based positioning. 

\subsubsection{Basics of Probability Maps}
We overlay a set of $K$ grid points $\bmg_k\in\reals^{D}$, $k=1,\ldots,K$, over the space that will be used to perform positioning. Here, $D$ is typically two or three and represents the number of spatial dimensions used to perform positioning, and the convex hull over all grid points 
\begin{align} \label{eq:convexhull}
	\setH = \left\{\sum_{k=1}^K \alpha_k \bmg_k \mid (\alpha_k\in\reals_+, \forall k) \wedge \sum_{k=1}^K \alpha_k = 1 \right\}
\end{align}
must include the target area in which localization will be performed. 
The probability map $\bmp^{(u)}[b]\in[0,1]^K$ represents the probability of UE $u$ being located at each grid point, i.e., we have that $p^{(u)}_k[b]\in[0,1]$ for $k=1,\ldots,K$ and $\sum_{k=1}^Kp^{(u)}_k[b]=1$. 
Note that the last softmax layer in the proposed NN structure shown in \fref{fig:nnmodel} ensures that the generated outputs correspond to  probability mass functions (PMFs).  
Furthermore, if the probabilities contained in $\bmp^{(u)}[b]$ indeed model the UE's position, we can compute the expected location of UE $u$ as follows:
\begin{align} \label{eq:expectedlocation}
	\hat\bmx^{(u)}[b] = \sum_{k=1}^K \bmg_k p^{(u)}_k[b].
\end{align}
By defining the $D\times K$ grid point matrix $\bG=\big[\bmg_1,\ldots,\bmg_K\big]$ the expected location is simply $\hat\bmx^{(u)}[b] = \bG \bmp^{(u)}[b]$.
Hence, one could easily augment the NN shown in \fref{fig:nnmodel} to directly generate an estimate of the UE location  $\hat\bmx^{(u)}[b] $  by adding an additional linear output layer with untrainable weights corresponding to $\bG$ and no bias terms. 

\begin{rem} 
	The selection of grid points can either form an equispaced rectangular grid for $D=2$ (or equispaced cubic grid for $D=3$) that includes the target area or can be chosen arbitrarily. An example of a rectangular grid is shown in \fref{fig:setup}. Irregular grid points may be useful to  only cover locations that are populated or to place grid points at higher density in areas where higher resolution is required. 
\end{rem}

\subsubsection{Training NNs with Probability Maps}
\label{sec:trainingofprobabilitymaps}
In order to ensure that the probability maps $\bmp^{(u)}[b]$ generated by the proposed NN accurately model the location of UE $u$ being at grid point $\bmg_k$ with probability $p^{(u)}_k[b]$, the network must be trained accordingly. 
Assume that we have a training set  with CSI feature vectors $\{\bmf^{(u)}[b]\}_{u=1}^{U'}$ obtained from $U'$ distinct locations $\{\bmx^{(u)}\}_{u=1}^{U'}$.
In order to train the NN shown in \fref{fig:nnmodel}, we need to compute reference probability maps $\{\bmp^{(u)}[b]\}_{u=1}^{U'}$ associated with ground-truth positions $\{\bmx^{(u)}\}_{u=1}^{U'}$. Unfortunately, given a position $\bmx^{(u)}$ there are, in general, infinitely many probability maps $\bmp^{(u)}[b]$ for which $\bmx^{(u)}[b]=\bG\bmp^{(u)}[b]$ holds and $\bmp^{(u)}[b]$ is a PMF. 
We propose to select the probability map for which the error variance is minimized---this only activates probabilities associated with grid points that are nearby the ground truth location.

To compute such minimum-variance probability maps, we reiterate that \fref{eq:expectedlocation} is nothing but the expected location of UE~$u$. The $D\times D$ covariance matrix is then given by
\begin{align} \label{eq:covariancematrix}
	\bC^{{u}}[b] = \sum_{k=1}^K  p^{(u)}_k[b] (\bmg_k - \hat\bmx^{(u)}[b]) (\bmg_k - \hat\bmx^{(u)}[b])^T.
\end{align}
and the combined variance is $\sigma^2[b] = \tr(\bC^{(u)}[b])$. By defining the vector $\bmv^{(u)}\in\reals^{K}$ with entries $v_k^{(u)}[b] =  \|\bmg_k- \hat\bmx^{(u)}[b]\|^2$, $k=1,\ldots,K$, we have that $\sigma^2[b] = \langle \bmv^{(u)}[b], \bmp^{(u)}[b]\rangle$. 
Hence, we can solve the following convex optimization problem to learn a minimum-variance probability map  $\bmp^{(u)}[b]$ from the ground-truth location~$\bmx^{(u)}$:
\begin{align} \label{eq:minvarianceprobabilitymap}
	\left\{
	\begin{array}{ll}
		\underset{\bmp\in\reals^{K}}{\text{minimize}}  &  \bmp^T \bmv^{(u)}[b]   \\
		\text{subject to}&   \| \bG\bmp -\x^{(u)}\|\leq \varepsilon, \\ &\sum_{k=1}^K  p_k =1,\,\, p_k \in [0,1], \forall k.
	\end{array}\right.
\end{align}
Here, $\varepsilon>0$ can be used to trade-off accuracy vs. variance.
Problems of this form can easily be solved using off-the-shelf convex solvers, such as CVX \cite{grant2014cvx}, or customized methods that build on Douglas-Rachford splitting \cite{douglasratchford}. Note that if a ground-truth position is outside the convex hull $\setH$ spanned by the grid points as defined in \fref{eq:convexhull}, then the optimization problem may no longer be feasible\footnote{Uniqueness depends on the choice of the trade-off parameter $\varepsilon$.}; for positions within the convex hull, the problem in \fref{eq:minvarianceprobabilitymap} is always feasible. 
For grid points on an equispaced 2-dimensional ($D=2$) grid, the problem in \fref{eq:minvarianceprobabilitymap} reduces to identifying the nearest four grid points (two in x-direction; two in y-direction) to the target location $\bmx^{(u)}$ and assign the probabilities to these four grid points while minimizing the variance.

After learning the minimum-variance probability maps  $\{\bmx^{(u)}\}_{u=1}^{U'}$ for each ground-truth position~$\bmx^{(u)}$, we can  learn the NN parameters $\bm{\theta}_b$. To this end, we use extracted CSI-features $\{\bmf^{(u)}[b]\}_{u=1}^{U'}$  and the probability maps $\{\bmp^{(u)}\}_{u=1}^{U'}$ associated with ground-truth position $\{\bmx^{(u)}\}_{u=1}^{U'}$, and we train the NN using a conventional binary cross-entropy loss function. 

\begin{rem}
	We have observed that using a cross-entropy loss instead of training the same network augmented with an additional grid point layer $\bG$ with a mean-square error loss between estimated and ground truth position resulted in superior positioning accuracy.  
\end{rem}

%%%%%
\section{Probability Fusion}
\label{sec:probabilityfusion}
As illustrated in \fref{fig:positioningpipeline}(b), we are interested in fusing multiple probability maps obtained from different APs and/or transmit antennas to improve positioning accuracy.
To explain the advantages of this approach, consider a situation in which two distinct locations yield similar CSI features at one AP, which would result in similar probability maps. While positioning from such probability maps would result in a large positioning error (as the two probability maps should describe two different locations), probability fusion can resolve such ambiguities. 
As long as the CSI acquired from another AP maps these two locations to \emph{distinct} probability maps, only the probabilities associated with locations where both maps agree would be strong after fusion---all other probabilities would be attenuated.
We now propose three different methods that implement probability fusion: Probability conflation, Gaussian conflation, and NN-based probability fusion. A comparison of these three fusion approaches is given in  \fref{sec:results}.

%%%
\subsection{Probability Conflation}
\label{sec:probabilityconflation}
Assume that we have multiple neural networks that generate different probability maps for a given UE~$u$, e.g., obtained from different APs $\bmp^{(u)}[b]$ , $b=1,\ldots,B$, or from different transmit antennas $\bmp^{(u)}_m[b]$, $m=1,\ldots,\MT$, or a combination of both. 
To simplify notation, assume that $B'$ neural networks generate a collection of $B'$ probability maps denoted by $\{\bmp^{(u)}[b]\}_{b=1}^{B'}$, irrespective of whether these are obtained from different APs or transmit antennas. 
Our goal is to fuse this collection of probability maps to a single probability map $\bar\bmp^{(u)}$, which can then be used to produce an improved estimate of the $u$th UE's position by computing the expected position $\hat\bmx^{(u)}=\bG \bar\bmp^{(u)}$. 

The idea of combining multiple PMFs into a single PMF that more accurately describes the observed quantity has been widely studied in the literature; see, e.g.,~\cite{hill2008conflations}.
Intuitively, combining PMFs for UE positioning should automatically give more weight to probability maps with smaller variance. Ideally, the  fused probability map should provide a more accurate estimate of the UE's position than solely using the most reliable probability map. 
To achieve this goal, we propose to use \emph{probability conflation} as put forward in \cite[Def.~2.7]{hill2008conflations}. 
The approach is straightforward---simply compute the (unnormalized) sub-PMF via a point-wise Hadamard product as
\begin{align} \label{eq:probconflation1}
	\mu^{(u)}_k = \prod_{b=1}^{B'}p^{(u)}_k[b], \quad k=1,\ldots,K,
\end{align}
followed by normalizing the fused sub-PMF vector $\boldsymbol\mu^{(u)}$ to the fused PMF according to
\begin{align}\label{eq:probconflation2}
	\bar\bmp^{(u)} = \frac{\boldsymbol\mu^{(u)}}{\|\boldsymbol\mu^{(u)}\|_1}.
\end{align}
As demonstrated in \cite[Sec.~4]{hill2008conflations}, probability conflation cannot improve the amount of information contained in all probability maps $\{\bmp^{(u)}[b]\}_{b=1}^{B'}$, but the resulting conflated PMF can be shown to be optimal (among other properties) in terms of minimizing the  loss of Shannon information.\footnote{According to \cite[Sec.~4]{hill2008conflations}, the Shannon information of an event $\setA$ is defined as $S(\setA)=-\log_2(P[\setA])$, where $P[\setA]$ is the probability of $\setA$.}

\begin{rem}
	Probability conflation requires one to pass all probability maps $\{\bmp^{(u)}[b]\}_{b=1}^{B'}$ with a total number of $K\times B'$ real numbers to a centralized processor, which calculates \fref{eq:probconflation1} and \fref{eq:probconflation2}. The complexity of probability conflation is $K(B'-1)+1+K\times D$ real-valued multiplications.
\end{rem}

%%%
\subsection{Gaussian Conflation}
\label{sec:gaussianconflation}
While probability conflation requires the transfer of $K\times B'$ real numbers, we can use an alternative conflation approach that reduces the amount of information transfer. 
This probability fusion approach is inspired by the method used to train probability maps in \fref{sec:trainingofprobabilitymaps}, where we compute the mean and variance of a probability map. 
Let us assume that the mean UE position $\hat\bmx^{(u)}$ can be modeled as follows:
\begin{align} \label{eq:Gaussianconflationmodel}
	\hat\bmx^{(u)}[b]  = \bmx^{(u)} + \bme^{(u)}[b].
\end{align}
Here, $\bmx^{(u)}[b]$ is the true position and the error vector $ \bme^{(u)}[b]\in\reals^D$ is assumed to be zero-mean. 
As shown in~\fref{eq:expectedlocation}, given a probability map~$\bmp^{(u)}[b]$, the mean position  can be computed as in \fref{eq:expectedlocation}.  
By furthermore assuming that the entries in the error vector $\bme^{(u)}[b]$ are pairwise uncorrelated (meaning that the positioning errors in each spatial dimension are uncorrelated), its covariance matrix $\bK^{(u)}[b]=\mathrm{diag}(\bC^{(u)}[b])$ corresponds to the  main diagonal of the covariance matrix $\bC^{(u)}[b]$ defined in~\fref{eq:covariancematrix}. 
With the model in  \fref{eq:Gaussianconflationmodel}, we have $B'$ ``noisy'' observations of the true location $\bmx^{(u)}$. By assuming that the entries in the error vector $\bme^{(u)}[b]$ are Gaussian and that the error vectors are pairwise independent across observations $b=1,\ldots,B'$, we can now perform \emph{Gaussian conflation} as analyzed in \cite[Thm.~6.1]{hill2008conflations}.
The optimal combination of mean positions $\hat\bmx^{(u)}[b]$, $b=1,\ldots,B'$, in terms of minimizing the post fusion error covariance (or mean-square error) is given by 
\begin{align} \label{eq:Gaussianconflation}
	\hat x^{(u)}_d = \frac{\sum_{b=1}^{B'} \big[\bK^{(u)}[b]\big]_{d,d}^{-1} \hat x_d^{(u)}[b]}{\sum_{b=1}^{B'} \big[\bK^{(u)}[b]\big]_{d,d}^{-1}  }, \quad d=1,\ldots,D.
\end{align}
Intuitively, Gaussian fusion de-weights position estimates with higher variance. 
The diagonal entries of the error covariance matrix~$\bK^{(u)}$ of the fused estimate $\hat \bmx^{(u)}_d $ from \fref{eq:Gaussianconflation} are given by 
\begin{align}
	\big[\bK^{(u)}\big]_{d,d} =  \sum_{b=1}^{B'} \big[\bK^{(u)}[b]\big]_{d,d}^{-1} , \quad d=1,\ldots,D.
\end{align}

\begin{rem}
	Gaussian conflation requires only $B'$ mean-variance pairs for each dimension $D$, which requires a transfer of $2\times D\times B'$ real numbers to a centralized processor, which computes an improved location estimate as in \fref{eq:Gaussianconflation}. The complexity of Gaussian conflation is $D\times B'\times K + D\times(2B' +1)$ real-valued multiplications. 
\end{rem}

%%%
\subsection{Neural-Network-Based Probability Fusion}
\label{sec:nnfusion}
Besides the two conflation methods discussed above, there exist other probability fusion methods. 
A straightforward approach is to compute simple unweighted average as
\begin{align} \label{eq:simpleaveraging}
	\hat\bmx^{(u)} = \frac{1}{B'} \sum_{b=1}^{B'} \hat\bmx^{(u)}[b] = \frac{1}{B'} \sum_{b=1}^{B'} \bG \bmp^{(u)}[b],
\end{align}
which is a special case of Gaussian conflation in \fref{eq:Gaussianconflation} assuming the error variances are equal. While this averaging approach can serve as a  baseline probability fusion method, it can be improved by a neural network that uses optimized linear combination weights. 

We first train the $B'$ neural networks $g_{{\boldsymbol\theta}_b}$, $b=1,\ldots,B'$. We then stack the $B'$ neural networks along with their individual probability map outputs $\bmp^{(u)}[b]$, $b=1,\ldots,B'$, and add a bias-free linear layer whose weight matrix is initialized with  
\begin{align} \label{eq:simpleaveragingmatrix}
	\bar\bG = \frac{1}{B'}\big[ \underbrace{\bG, \ldots, \bG}_{B'\text{ times}} \big]
\end{align}
to its output. This final linear layer, combined with the stacked probability map vector 
\begin{align}
	\bar\bmp^{(u)} = \big[ \bmp^{(u)}[1]^T ,\ldots,  \bmp^{(u)}[B']^T \big]^T,
\end{align}
computes $\hat\bmx^{(u)} =\bar\bG  \bar\bmp^{(u)} $ as in  \fref{eq:simpleaveraging}. 
For the same training set consisting of ground-truth locations $\{\bmx^{(u)}[b]\}_{u=1}^{U'}$  and associated CSI-feature vectors $\{\bmf^{(u)}[b]\}_{u=1}^{U'}$, we then continue weight learning of the final layer $\bar\bG$ by minimizing the mean distance error (MDE) loss  defined as
\begin{align} \label{eq:MDE}
	L_\text{MDE} = \frac{1}{U'}\sum_{u=1}^{U'} \big\|\bmx^{(u)}-\hat\bmx^{(u)}\big\|_2.
\end{align}
Since we continue learning $\bar\bG$ after initializing it with the matrix in \fref{eq:simpleaveragingmatrix}, we expect this method to perform no worse than keeping $\bar\bG$ fixed and performing averaging as in \fref{eq:simpleaveraging}. 

\begin{rem}
	NN-based probability fusion requires one to pass all probability map estimates with a total number of $K \times B'$ real numbers to a centralized processor, which multiplies the estimates with the trained weights. 
	The complexity of NN-based probability fusion is $D\times B' \times K$ real-valued multiplications. 
\end{rem}

\begin{rem}
	One could also retrain the weights and biases contained in the network parameters~$\boldsymbol\theta_b$, $b=1,\ldots,B$, of the $B'$ networks when learning the weights in the matrix $\bar\bG$. We have, however, not observed accuracy improvements by doing so.
\end{rem}

\section{Results}
\label{sec:results}

We now evaluate the performance of the proposed positioning pipeline for a range of  indoor CSI measurements. We first describe the system setup, measurement scenarios, and performance metrics. 
We then show accuracy results for a range of multi-antenna and multi-point probability fusion methods.

\subsection{Measurement Setup}
\label{sec:measurementsetup}

\begin{figure}[tp]
	\centering
	\includegraphics[width=0.3\textwidth]{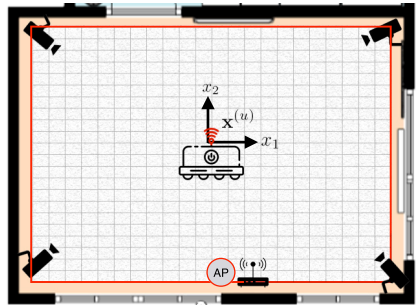}
	\caption{Measurement setup. We use a robot that carries a WLAN transmitter; one or multiple APs then record CSI measurements. An accurate position estimate of the robot $\bmx^{(u)}$ is extracted using a multi-camera precision localization system and recorded for NN training and testing.}
	\label{fig:setup}	
\end{figure}

Our measurement setup consists of five components: 
(i) A portable UE with wireless LAN (WLAN) connectivity; we use a Raspberry Pi Model 4 that is equipped with a two-antenna IEEE 802.11ac transceiver. 
(ii) A robot equipped with an embedded processor to move the UE; we use iRobot Roomba Create 2 controlled by the Raspberry Pi. 
(iii) A WLAN AP that enables the extraction of CSI; the AP is equipped with a four-antenna IEEE 802.11ac transceiver 
operating at 5\,GHz with 80\,MHz bandwidth and provides an API to access the raw per-subcarrier CSI for each receive and transmit antenna. 
(iv) A precise positioning system for ground-truth position extraction; we use use two different systems depending on the scenario: A WorldViz PPT-N active point tracking system \cite{worldviz} with sub-millimeter positioning accuracy enabled by triangulation with four infrared (IR) camera readings of active IR transmitters, and Vicon Vero passive point tracking system \cite{vicon} with sub-millimeter positioning accuracy enabled by triangulation with twelve motion capture camera readings of passive reflective markers.
(iv) A host computer that is collecting CSI measurements, running the precision positioning systems, and controlling the robot. 
\fref{fig:setup} illustrates the measurement setup.

The data collection procedure is as follows: 
We control the robot's position to follow a predefined path in piecewise linear movements over two dimensions $\bmx=[x_1,x_2]^T$. 
The Raspberry Pi continuously transmits high-quality images from an on-board camera to the host PC via the two transmit antennas $\MT=2$.
At the same time, the AP is receiving data and CSI from the four antennas $\MR=4$, and the precision positioning system is extracting ground-truth position information; all the data is stored at the host computer. 
Since we measure CSI and ground truth positions from two separate sources, we first synchronize the operating-system clocks of the Raspberry Pi and the precision positioning system running at the host PC via the network time protocol (NTP). We then match the measurement times of the CSI and the precision positioning system.
For the multi-point measurement, we perform two separate measurement campaigns, i.e., we first measure CSI at AP1 and ground-truth position for the entire area and later we measure CSI at AP2 and ground-truth position while following approximately the same track with the robot. We then match the CSI from both APs so that there is less than $3$\,cm ground-truth position difference between both measurements.

\begin{rem}
	There exists a trade-off between the training dataset size and the positioning accuracy.
	We have observed that larger training sets generally improve accuracy (with diminishing returns), but also increase the time to collect such datasets.
	The optimal amount of training required in practice depends on the capabilities of the positioning system (in terms of minimum achievable positioning error), the scenario, as well as the time required to acquire the dataset.
\end{rem} 

\begin{rem}
	Since our multi-AP measurements are acquired at two different time instants---more than tens of minutes apart---our results  in \fref{sec:positioningresults} imply that accurate positioning from multi-AP measurements does not need exact synchronization or simultaneous recording of CSI.
\end{rem}

\begin{figure}[tp]
	\centering
	\subfloat[]
	{	\centering
		\includegraphics[width=0.22\textwidth]{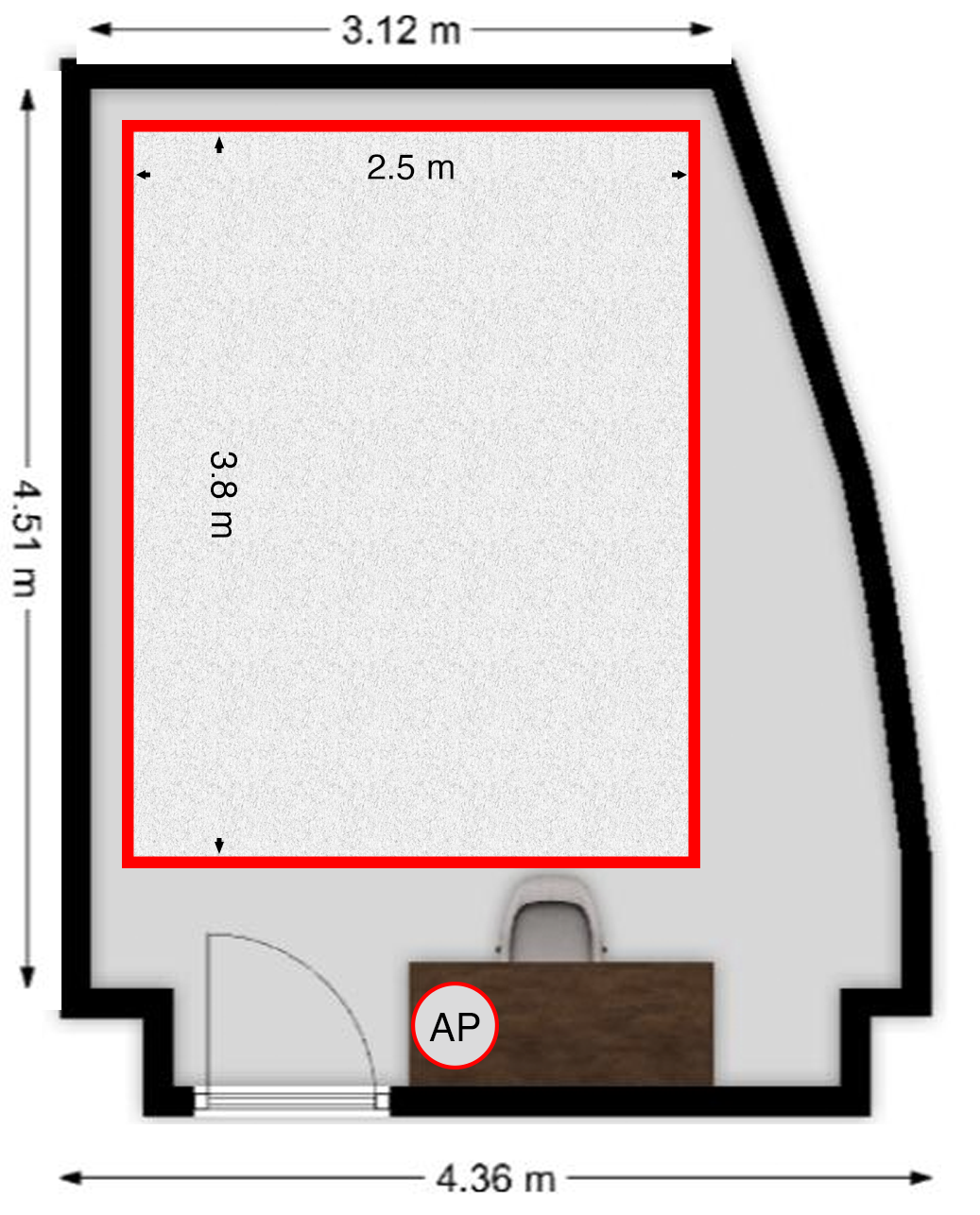}
	}\hspace{0.2cm}
	\subfloat[]
	{\centering
		\includegraphics[width=0.22\textwidth]{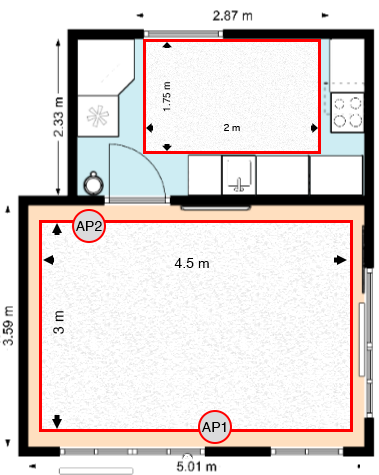}
	}
	\caption{Floor plans used to perform indoor positioning experiments. (a) Lab floor plan with an area of $2.5\text{m}\times3.8\text{m}$.  (b) Living room (bottom; area $3\text{m}\times4.5\text{m}$)   and kitchen (top; area $1.75\text{m}\times2\text{m}$) floor plans. The access point (AP) locations are marked by circles.}
	\label{fig:floorplans}
\end{figure}

%%%
\subsection{Measurement Scenarios}
\label{sec:measurementscenarios}
In order to evaluate the proposed multi-antenna and multi-point probability fusion positioning pipeline, we consider four different scenarios that were measured in three different locations: Lab space, living room, and kitchen.
\fref{fig:floorplans} shows the floor plans of these locations as well as the AP positions.
For each location, we divided the data acquisition process into two phases: In the first phase, we recorded a training set containing CSI measurements and ground-truth locations.
In the second phase, we recorded a separate test set that is used solely for characterizing the position accuracy of our positioning pipeline.\footnote{We note that CSI-based positioning methods that utilize DNNs do not generalize well to untrained areas; see, e.g.,  \cite{tenbrink2018generalizationcsi}.}
During both of these phases, human activity was present in the environment, including people walking around the room, opening and closing doors, turning small appliances off and on, etc. 
This data collection approach prevents overfitting and demonstrates a certain degree of robustness to dynamic changes in the environment.\footnote{We note that it is challenging to acquire a dataset in which the dynamic changes in the environment are well-controlled and reproducible.}
%

%%%
\subsubsection*{a) Single-Point LoS Lab Scenario}
\label{sec:loslab}
We collected CSI and ground-truth position in the lab space shown in \fref{fig:floorplans}(a) under line-of-sight (LoS) conditions. 
For the training set, the robot moved through a pre-defined area in the lab space over two days; \fref{fig:scenarios}(a) shows the robot's path. For the test set, the robot followed a ``VIP'' shaped path (shown with blue color) which was recorded separately from the training data on  another day.

%%%	
\subsubsection*{b) Single-Point Non-LoS Living Room Scenario}
We collected CSI and ground-truth position in the living room shown at the bottom of \fref{fig:floorplans}(b) under non-LoS conditions; non-LoS conditions are achieved by placing AP1 behind a TV.
For the training set, the robot was moving randomly through a predefined area in the living room; \fref{fig:scenarios}(b) shows the robot's path.
For the test set, the robot followed a ``VIP'' shaped path (shown with blue color), which was recorded separately but on the same day.

%%%	
\subsubsection*{c) Single-Point Non-LoS Kitchen Scenario}
We collected CSI and ground-truth position in the kitchen shown at the top of  \fref{fig:floorplans}(b) under non-LoS conditions as we used AP1.
For the training set, the robot was moving randomly through a predefined area in the kitchen;  \fref{fig:scenarios}(c)  shows the robot's path. 
For the test set, we used $20$\% randomly selected measurements from the robot's path (shown with blue color), which were recorded separately but on the same day.

%%%	
\subsubsection*{d) Multi-Point LoS Living Room Scenario}
We collected CSI and ground-truth position in the living room area shown at the bottom of  \fref{fig:floorplans}(b) under LoS conditions, where we used both AP1 and AP2. 
For the training set, we recorded CSI at two distinct times on the same day: one recording for AP1 and one for AP2. For both of these recordings, the robot was following a pre-defined path; \fref{fig:scenarios}(d) shows the robot's path.
For the test set, the robot followed a rectangular-shaped path (shown with blue color), which was recorded separately but on the same day.

\newcommand{\figsixsize}{0.22}
\newcommand{\figsixspace}{0.1cm}
\begin{figure*}[tp]
	\center
	\label{fig:scenario}	
	\subfloat[]
	{
		\includegraphics[width=\figsixsize\textwidth]{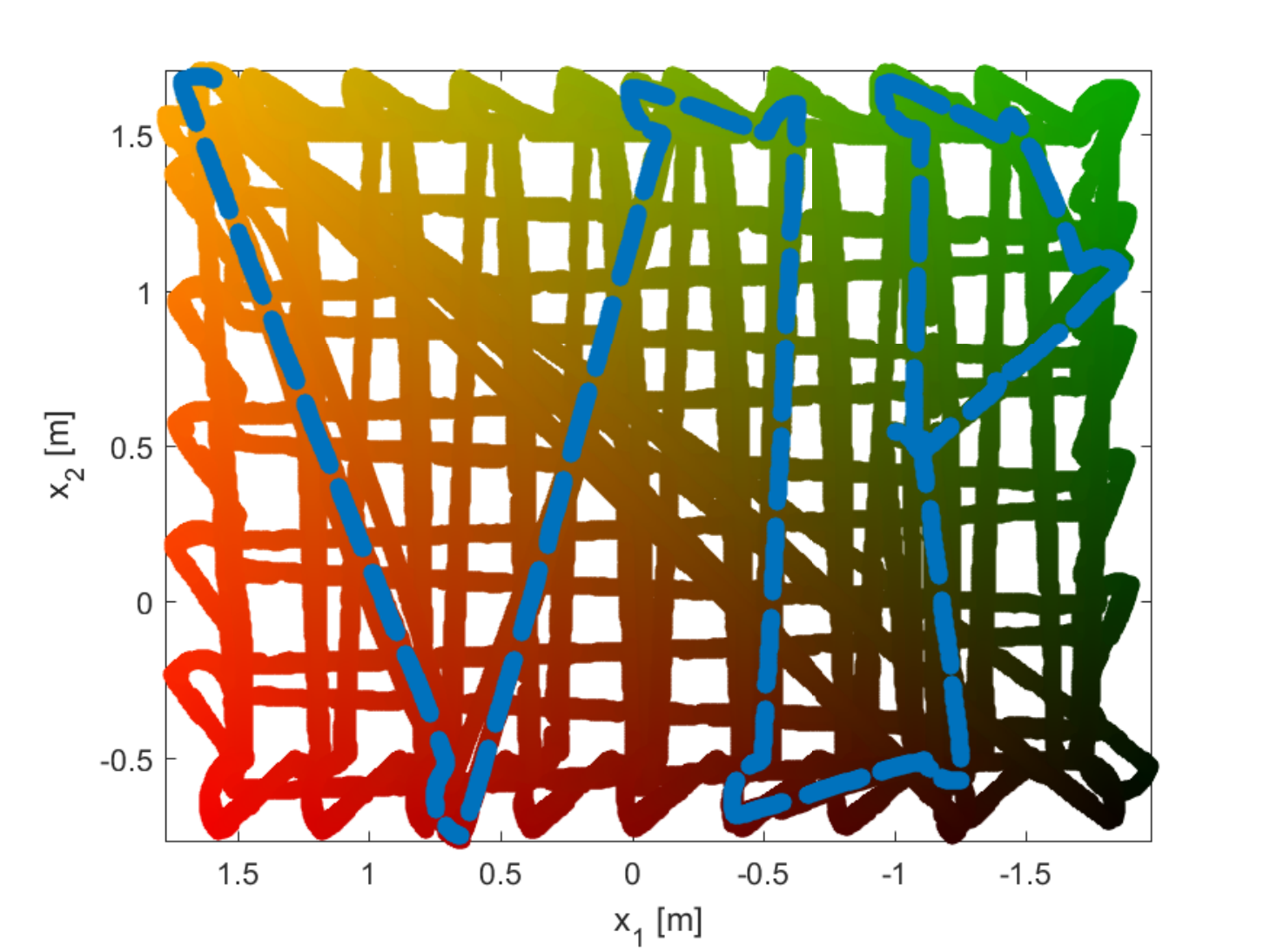}
	}\hspace{\figsixspace}
	\subfloat[]
	{
		\includegraphics[width=\figsixsize\textwidth]{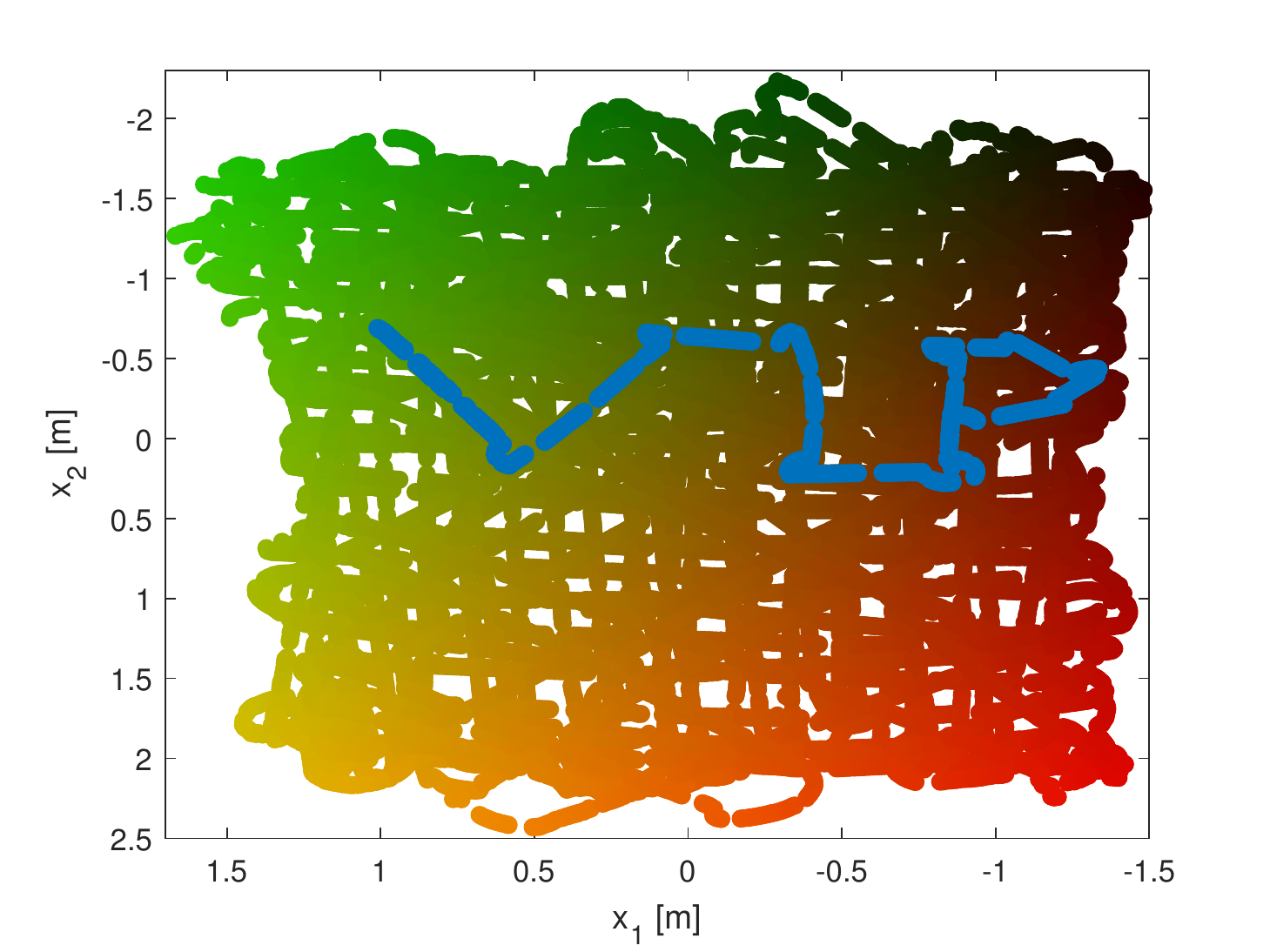}
		
	}\hspace{\figsixspace}
	\subfloat[]
	{		
		\includegraphics[width=\figsixsize\textwidth]{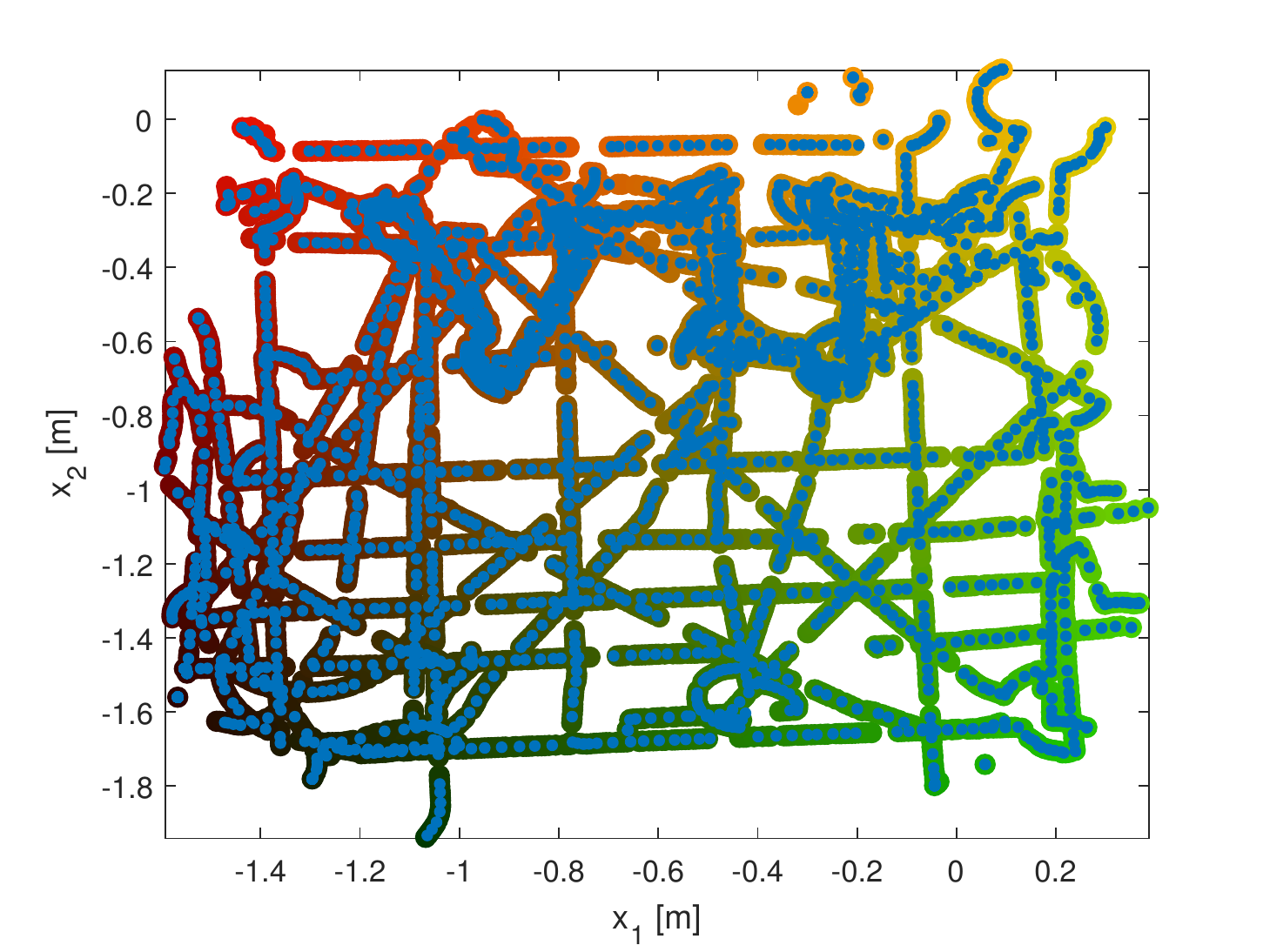}				
	}\hspace{\figsixspace}
	\subfloat[]
	{		
		\includegraphics[width=\figsixsize\textwidth]{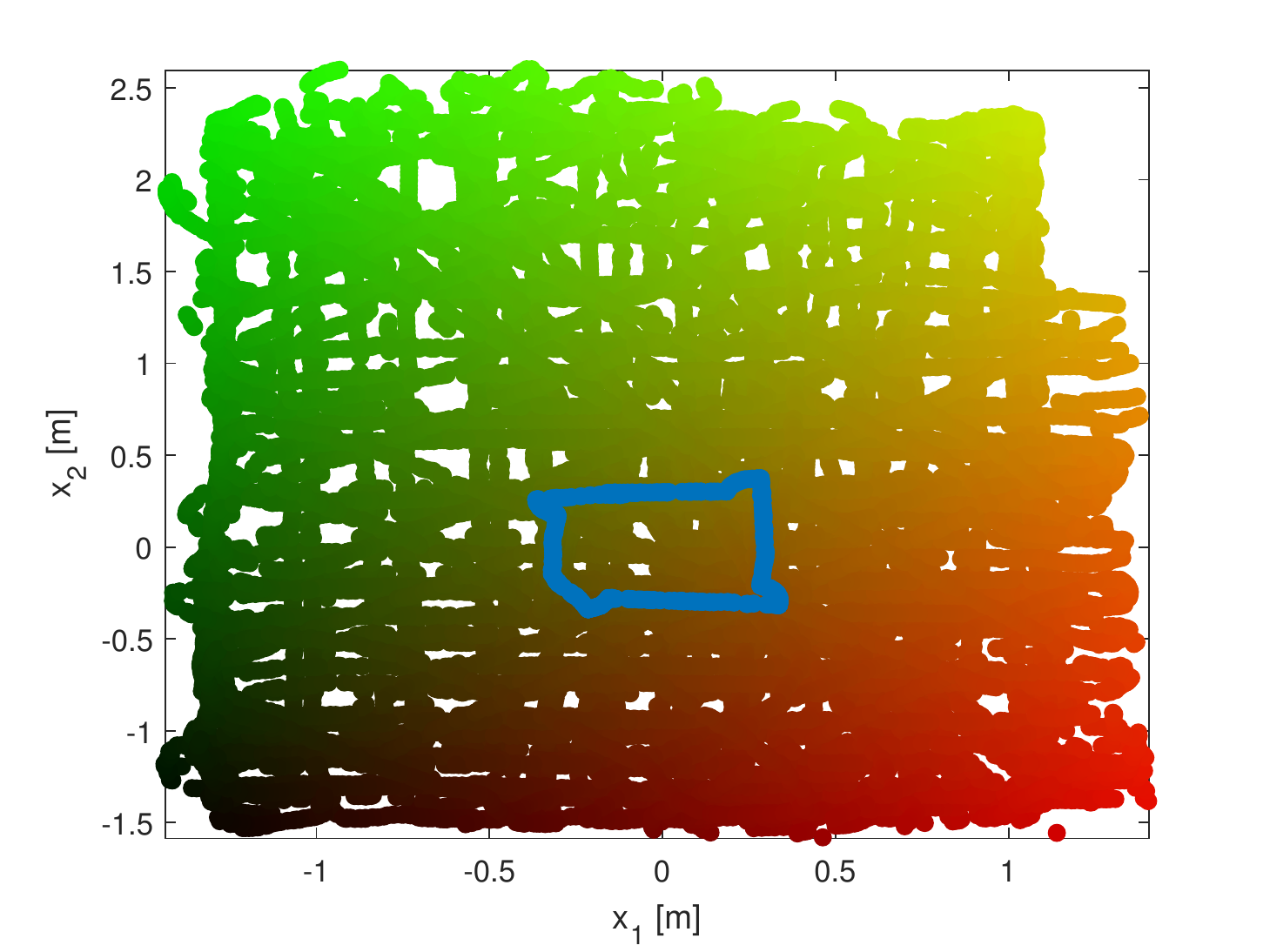}
	}
	\caption{Ground-truth positions collected for different scenarios: (a) Single-point LoS lab; (b) single-point non-LoS living room; (c) single-point non-LoS kitchen; (d) multi-point LoS living room. The gradient-colored curves represent locations used for training; the blue curves correspond to the test set. }
	\label{fig:scenarios}
\end{figure*}

\subsection{Performance Metrics and Positioning Methods}

\begin{figure}[tp]
	\centering
	\includegraphics[width=0.7\columnwidth]{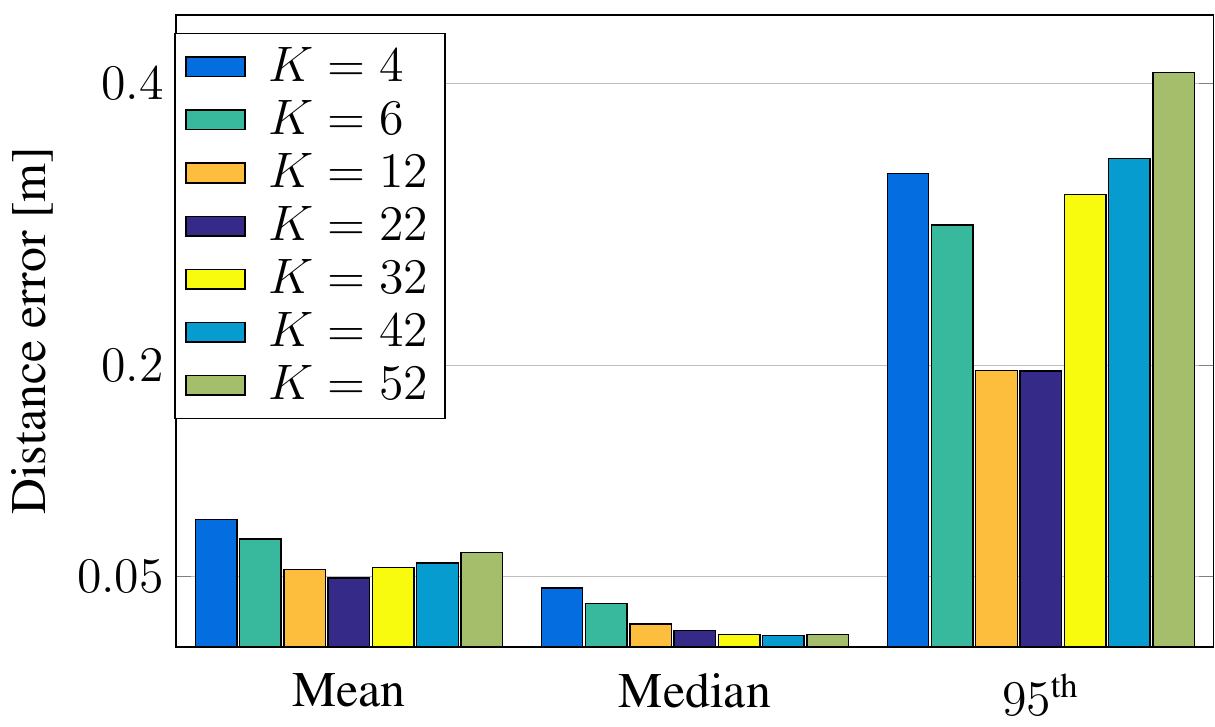}
	\caption{Impact of grid resolution on distance error. A $K\times K$ grid with $K=22$ yields the lowest mean and 95th percentile errors. Increasing the resolution further ($K>22$) does not provide significant improvements in terms of the median distance error.}
	\label{fig:gridsize}
\end{figure}

For all experiments, we use the CSI-features detailed in \fref{sec:CSIfeatures} and the NN topology discussed in \fref{sec:nnstructure}.
In order to determine the optimal grid resolution, we have conducted experiments with different $K\times K$ rectangular uniform grids with $K \in \{4, 6, 12, 22, 32, 42\}$ for the single-point LoS lab scenario.
\fref{fig:gridsize} shows the associated distance errors. We see that a resolution of $K = 22$ (which leads to a total of $K^2 = 484$ grid points) results in the lowest mean and 95th percentile errors, which is what we have selected for all of our experiments. 

In order to evaluate the positioning performance of the compared methods, we first obtain position estimates for the test set and then we compute three different metrics: (i) The mean distance error (MDE), i.e., the expression in \fref{eq:MDE} evaluated over the test set, (ii) the median distance error, and  (iii) the 95th percentile error. The median and 95th percentile distance errors were obtained by sorting the distances $\|\bmx^{(u)}-\hat\bmx^{(u)}\|_2$, $u=1,\ldots,U'$, of the test set in ascending order and extracting the 50\% and 95\% distances, respectively. 

For the multi-antenna and multi-point scenarios, we use the probability fusion methods discussed in \fref{sec:probabilityfusion}.
Concretely, we compare the performance of CSI-based positioning for the following methods.
For single-point and single-antenna experiments, we train a single neural network (NN) for one transmit antenna and one AP; we label these methods as ``1 NN, AP$b$ TX$m$,'' where ``1 NN'' implies that we use a single NN, $b=1,\ldots,B$ is the AP number, and $m=1,\ldots,\MT$ the transmit (TX) antenna number. 
For multi-antenna and/or multi-point experiments, we also  a single NN that stacks all features extracted from all $B$ APs in a single long CSI-feature vector\footnote{We note that the same approach has been proposed in \cite{li2020wireless}, which appeared after the submission of our manuscript.}, i.e., $\bmf^{(u)}_m[b]$ for  $b=1,\ldots,B$ and $m=1,\ldots,\MT$.
We label this method as ``1 NN, stacked features.''

In what follows, we will compare this method to probability fusion-based approaches. 
For fusion-based methods, we train two or four NNs depending on the scenario.
For two transmit antennas and one AP (i.e., multi-antenna and single-point) or for two APs and one transmit antenna (i.e., single-antenna and multi-point), we train two NNs and perform probability fusion from two probability maps. 
We consider the following methods: 
``2 NN, averaging,'' where we   implement unweighted averaging as in \fref{eq:simpleaveraging}; 
``2 NN, prob.~conflation,'' where we implement probability conflation as detailed in \fref{sec:probabilityconflation};  
``2 NN, Gaussian conflation,'' where we implement Gaussian conflation as detailed in \fref{sec:gaussianconflation}; and 
``2 NN, NN fusion,'' where we implement the NN-based fusion as detailed in \fref{sec:nnfusion}.
For two transmit antennas and two APs (i.e., multi-antenna and multi-point), we train four NNs and perform probability fusion from four probability maps; the associated methods are labeled as ``4 NN, averaging,'' ``4 NN, prob.~conflation,'' ``4 NN, Gaussian conflation,'' and ``4 NN, NN fusion.''

\begin{figure}[tp]
	\center
	\includegraphics[width=0.7\columnwidth]{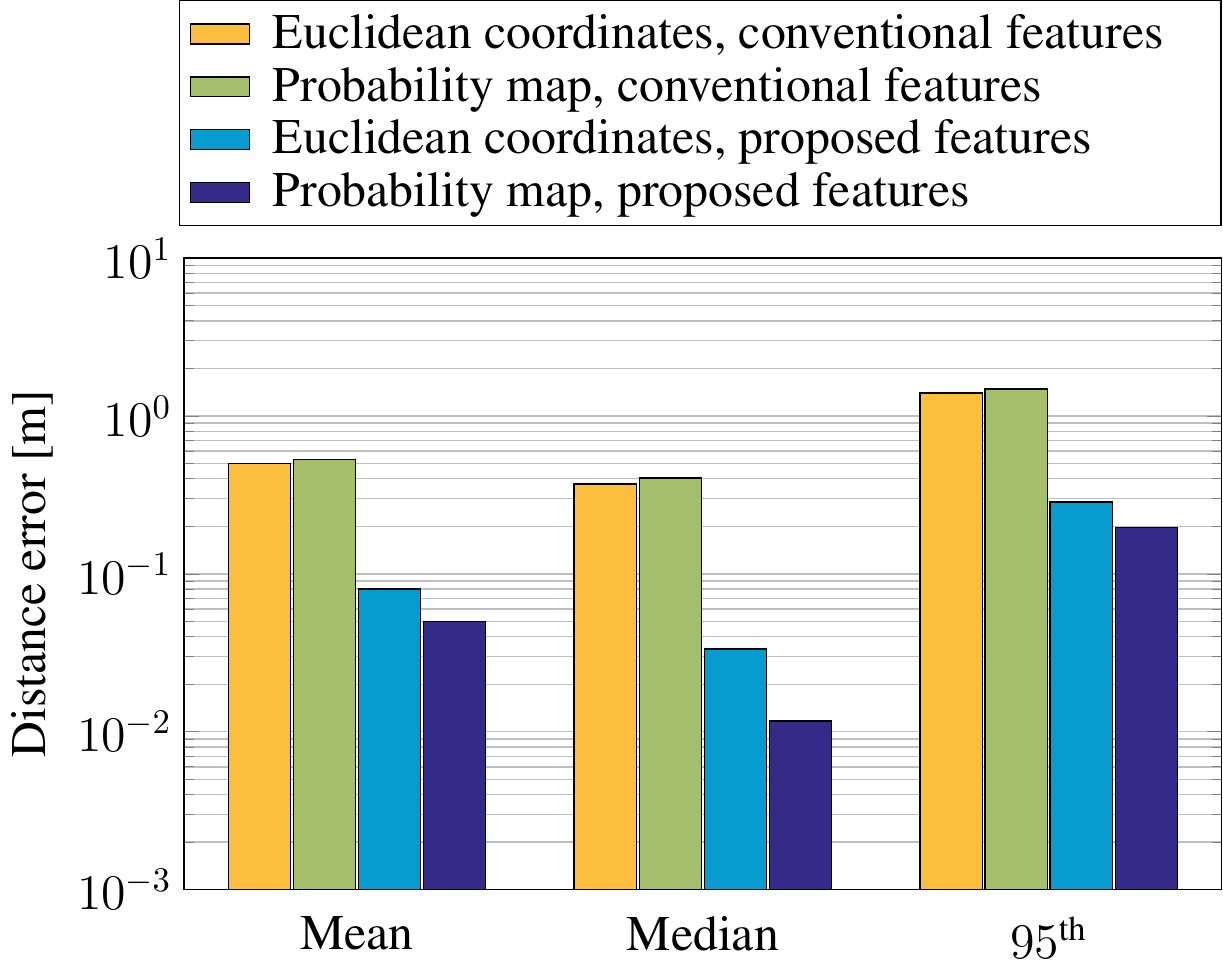}
	\caption{Bar plot showing mean, median, and 95th percentile distance error evaluated on the test set for the single-point LoS lab scenario detailed in \fref{sec:measurementscenarios}. We show a NN trained with the CSI features from \cite{ericpaper} using a NN that computes 2D coordinates (orange), and probability maps (green); and with features as in \fref{sec:CSIfeatures} using (i) a NN that computes 2D coordinates (light-blue), and (ii) a NN that computes probability maps (dark-blue).	Combining our CSI features with probability maps results in significant accuracy improvements.}
	\label{fig:compare}
\end{figure}

\subsection{Comparison with Single-AP Localization Approaches}

Before we discuss the experimental results for multi-AP positioning, we first assess the impact of CSI features and probability maps on positioning accuracy on a single-AP scenario. 
We consider the single-point LoS scenarios  in \fref{sec:loslab} with \textit{conventional} CSI features proposed in \cite{ericpaper} which uses magnitudes of beamspace representation of CSI measurements as features. 
Figure \ref{fig:compare} depicts the mean, median, and 95th percentile distance error obtained by (i) training a NN that generates Euclidean coordinates using conventional CSI features \cite{lundpaper,ericpaper}\footnote{Reference \cite{lundpaper} uses convolutional neural networks (CNNs) to estimate location. In our setup, we have observed no improvement in using deep CNNs over simple feedforward networks.} discussed in \fref{sec:beamspace_delaydomain_transform}
(orange), (ii) training a NN that generates probability maps as in \fref{sec:probabilitymaps} using the CSI features as in~\cite{lundpaper,ericpaper}  (green), (iii) training a NN which generates Euclidean coordinates using our feature extraction approach (light-blue) as in \fref{sec:autocorrelation}, and (iv) training a NN that generates probability maps as in \fref{sec:probabilitymaps} using our feature extraction approach (dark-blue) as in \fref{sec:autocorrelation}.
For the NN that generates Euclidean coordinates, we use the same architecture as for the NN that generates probability maps described in \fref{sec:nnstructure}, and we use a mean-squared-error loss function.
We observe significant accuracy improvements with our CSI features for real-world 802.11ac-OFDM measurements over conventional features that have been designed specifically for outdoor multi-user (MU) massive MIMO-OFDM channels. 
In addition, the use of probability maps not only results in an additional $3$\,cm MDE improvement, but also enables probability fusion from multiple APs as detailed in \fref{sec:probabilityfusion}.

\begin{rem}
	We have conducted \emph{synthetic} experiments with the same positioning pipeline where we used the positions obtained in the single-AP scenario to generate CSI using the QuaDRiGa ``WINNER\_Indoor\_A1\_LOS'' channel model~\cite{quadriga} with the same system parameters.
	With our positioning pipeline, we have observed a $0.9$\,cm mean distance error, a $0.8$\,cm median distance error, and a $2.3$\,cm $95$th percentile error.
	However, such synthetic experiments are not representative as they lack of real-world propagation conditions and hardware impairments.
\end{rem}

\subsection{Positioning Results}
\label{sec:positioningresults}

\fref{fig:barmultiantenna} shows bar plots evaluated on the test sets for multi-antenna  scenarios. 
\fref{fig:barmultipoint} shows bar plots evaluated for a multi-point scenario. 
\fref{fig:barmultipointmultiantenna} shows bar plots for a multi-antenna multi-point scenario. 
For all three figures, the left bar plot shows the mean distance error, the middle bar plot shows the median distance error, and the right bar plot shows the 95th percentile distance error. 

\subsubsection{Multi-Antenna Results}
Figures \ref{fig:barmultiantenna}(a), \ref{fig:barmultiantenna}(b), and \ref{fig:barmultiantenna}(c) show positioning results corresponding to the multi-antenna scenarios a), b), and c) detailed in \fref{sec:measurementscenarios}.
We observe that when using CSI-features from different transmit antennas but from a single AP, the accuracy can vary significantly across antennas. 
For the probability fusion methods, we see that simple averaging and NN-based fusion performs equally well.
The best performing methods are the use of a single NN with stacked features (labeled by ``1 NN, stacked features''), as well as probability conflation (labeled by ``2 NN, prob. conflation'') and Gaussian conflation (labeled by ``2 NN, Gaussian conflation''). 

\subsubsection{Multi-Point Results}
\fref{fig:barmultipoint} shows positioning results corresponding to the multi-point scenario d) detailed in \fref{sec:measurementscenarios} where we only use one transmit antenna. 
We observe that the accuracy of using AP1 is superior than that of AP2. 
Considering the APs have identical hardware, we conclude that the AP position may affect the positioning accuracy.
Furthermore, we see that when fusing the probability maps from AP1 and AP2, probability conflation and Gaussian conflation performs equally well as the single NN with stacked features. However, the amount of data to be transferred to a centralized processor for the 
Gaussian conflation approach is significantly lower than those of the single NN with stacked features and probability conflation.

\subsubsection{Multi-Antenna Multi-Point Results}
\fref{fig:barmultipointmultiantenna} shows positioning corresponding to the multi-point scenario detailed in \fref{sec:measurementscenarios} where we use both transmit antennas.
We see that the accuracy of the first transmit antenna (TX1) of AP1 is superior to the other antenna-AP combinations. 
We also see that the AP position and the used transmit antennas have an effect on the positioning accuracy.
Furthermore, we see that Gaussian conflation outperforms all other fusion approaches for the considered performance metrics.

\subsubsection{Visualization of Probability Fusion}
\fref{fig:multipointfusionresults} illustrates the efficacy of probability fusion, where we show the ground-truth positions in \fref{fig:multipointfusionresults}(a) on the test-set for scenario d) in \fref{sec:measurementscenarios} and estimated position using the proposed methods. 
\fref{fig:multipointfusionresults}(b) shows that the estimated locations at AP2 from the second transmit antenna (TX2) results in quite a few outliers.
When performing a single NN with stacked features and multiple NNs with Gaussian conflation in \fref{fig:multipointfusionresults}(c) in \fref{fig:multipointfusionresults}(d), respectively, one can see that the position accuracy significantly improves. 
We observe that probability fusion yields superior accuracy than using only the individual AP estimates. Furthermore, we see that NN-based probability fusion performs on par with simple averaging method. We speculate that much more training data and more complicated neural networks would be required to improve NN-based probability fusion.
Clearly, treating the information contained in the probability maps as Gaussians and fusing these mean-variance pairs results in accurate indoor positioning while only requiring a minimum amount of data transmission to a centralized processor that performs fusion. 
Among the studied probability fusion methods, we note that each method has pros and cons in terms of  complexity and data transfer size. For a use-case where the data transfer and/or complexity is not a concern, one could use either probability conflation or NN-based probability fusion. For a use-case where data transfer and/or complexity is severely constrained, Gaussian conflation often outperforms the other probability fusion methods.

\begin{rem}
	We emphasize that the 1 NN method with a single stacked feature results in excessively large features in the first layer, which substantially increases computational complexity and storage. Furthermore, the stacked feature approach requires centralized processing of all CSI features, which requires a large number of data to  be transferred to a centralized processor. In contrast, Gaussian conflation requires the transfer of only mean-variance pairs to the centralized processor, while resulting in similar or often superior accuracy. 
\end{rem}

\begin{rem}
	We have identified six key factors that affect positioning accuracy: (i) line-of-sight connectivity, (ii) the number of transmit and receive antennas, (iii) the number of APs, (iv) the number of subcarriers, (v) the number of training locations, and (vi) the CSI features.
	We expect to achieve superior accuracy in both LoS and NLoS conditions with more APs and with more receive antennas. 
	Moreover, we have observed that larger training sets reduce positioning accuracy; however, the acquisition of large datasets is not always feasible due to practical limits (e.g., acquisition time or storage).
	We note that the positioning accuracy is likely to degrade for areas that were not included in the training set~\cite{tenbrink2018generalizationcsi}.
	We hope to be able to reduce the amount of training and to improve generalization to untrained areas by combining our pipeline with self-supervised localization methods, such as channel-charting~\cite{channelcharting}.
\end{rem}

\newcommand{\figsize}{0.28}
\newcommand{\figsizef}{0.25}
\newcommand{\figsizekf}{0.245}
\newcommand{\fighspace}{1.4cm}
\begin{figure*}[tp]
	\center
	\begin{minipage}{\textwidth}
		\centering
		~~\includegraphics[width=0.915\textwidth]{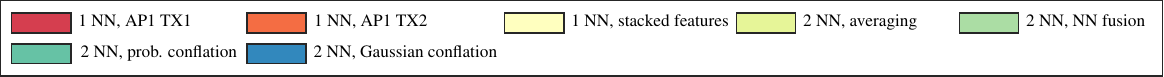}
	\end{minipage}\\[0.2cm]
	\subfloat[]{
		\includegraphics[width=\figsizef\textwidth]{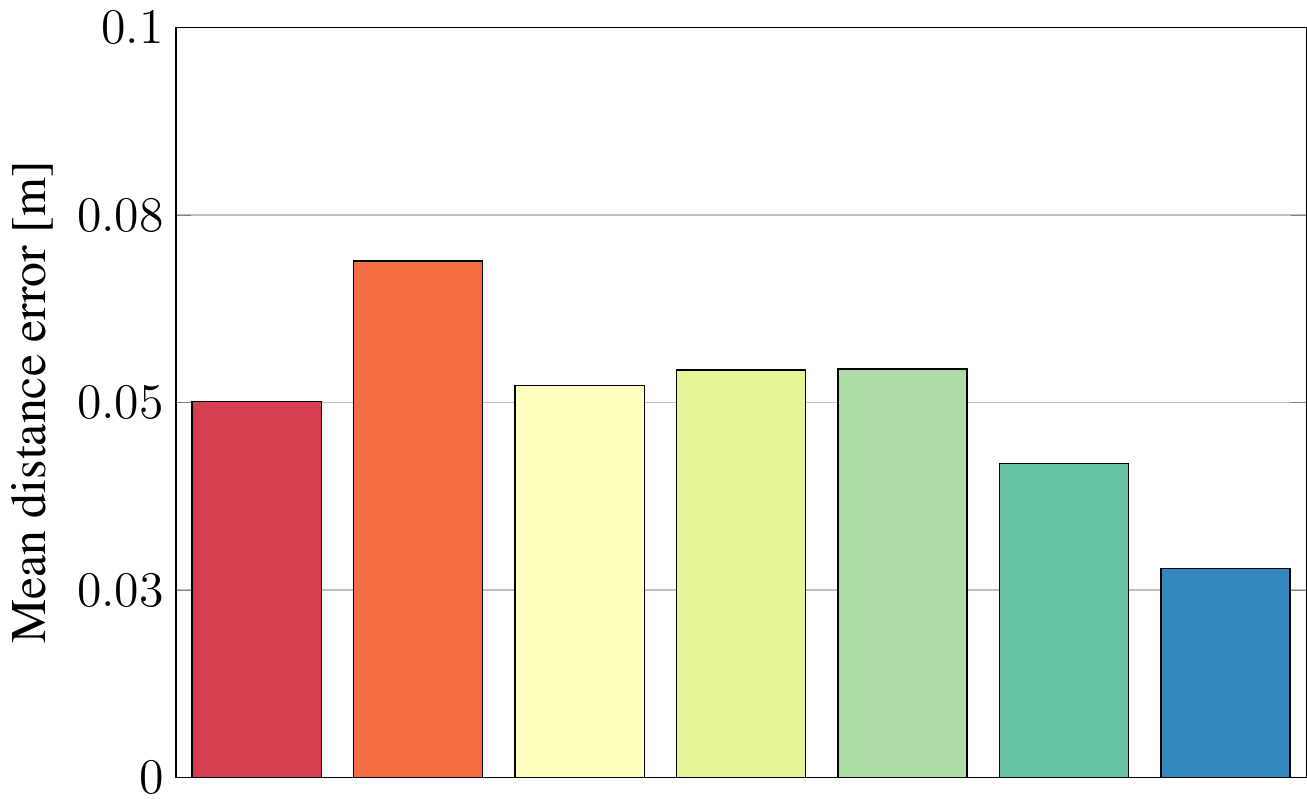}
		\hspace{\fighspace}
		\includegraphics[width=\figsizef\textwidth]{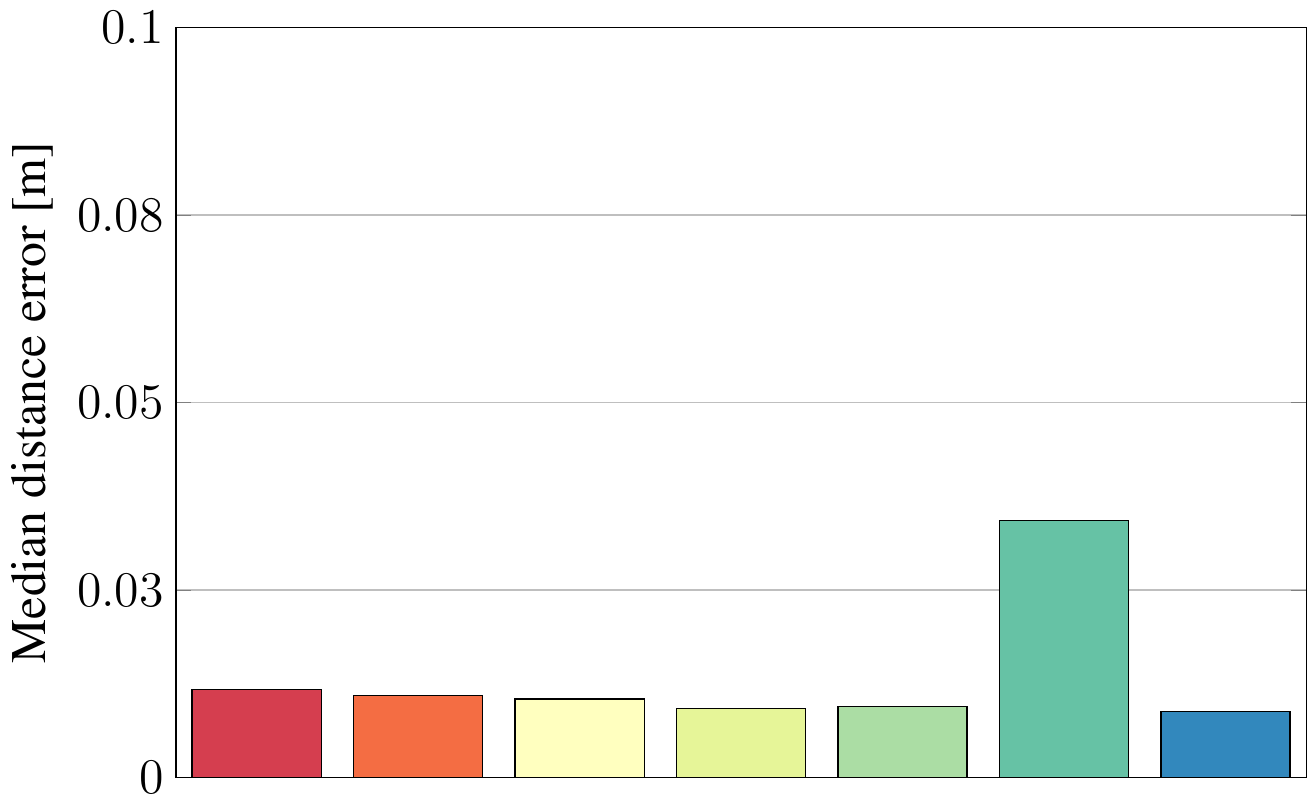}
		\hspace{\fighspace}
		\includegraphics[width=\figsizekf\textwidth]{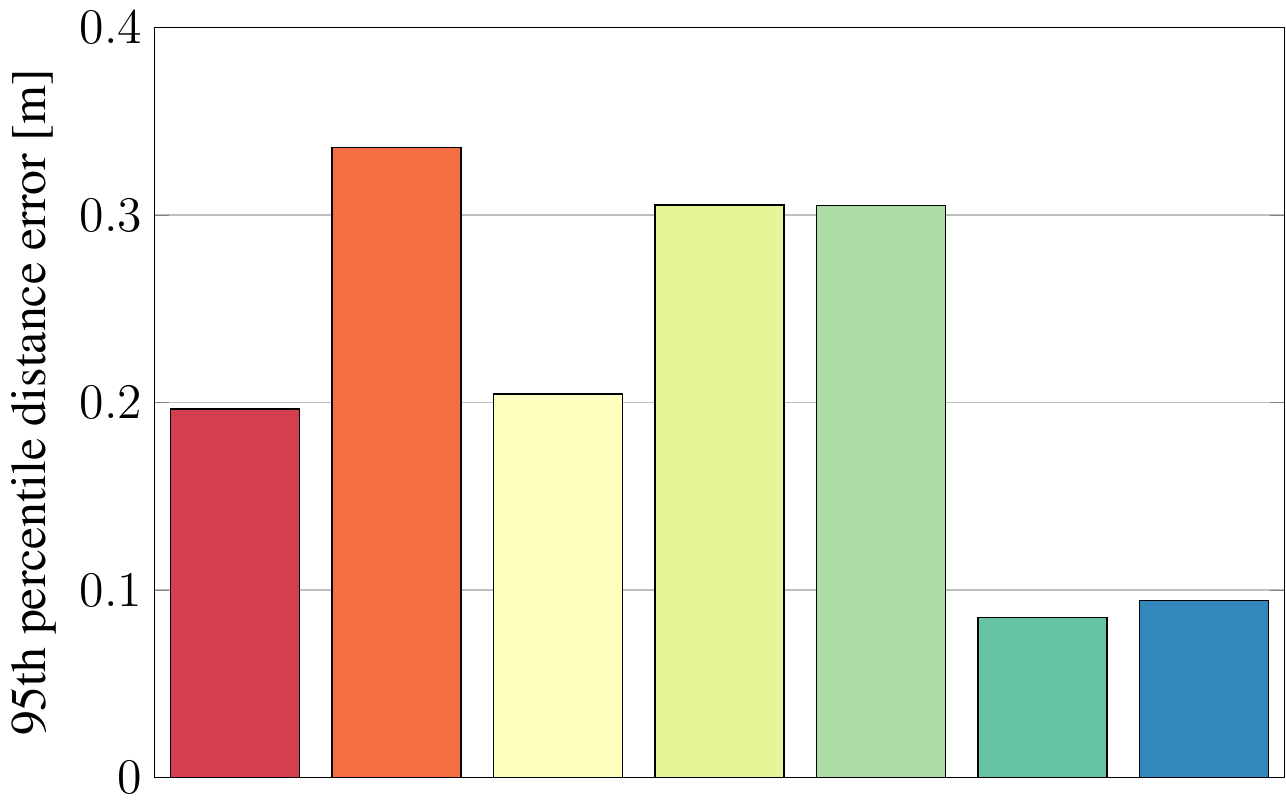}
	}
	\\
	\subfloat[]
	{
		\includegraphics[width=\figsizef\textwidth]{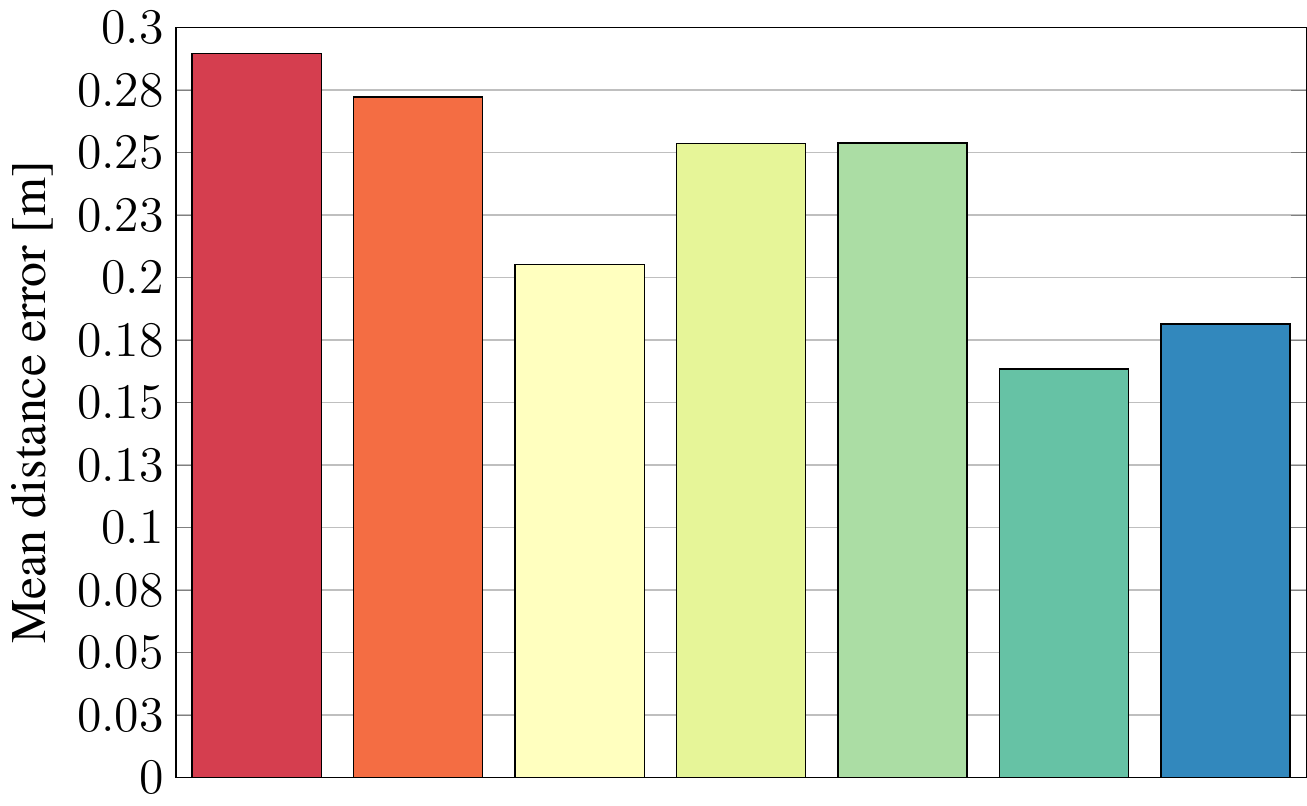}
		\hspace{\fighspace}
		\includegraphics[width=\figsizef\textwidth]{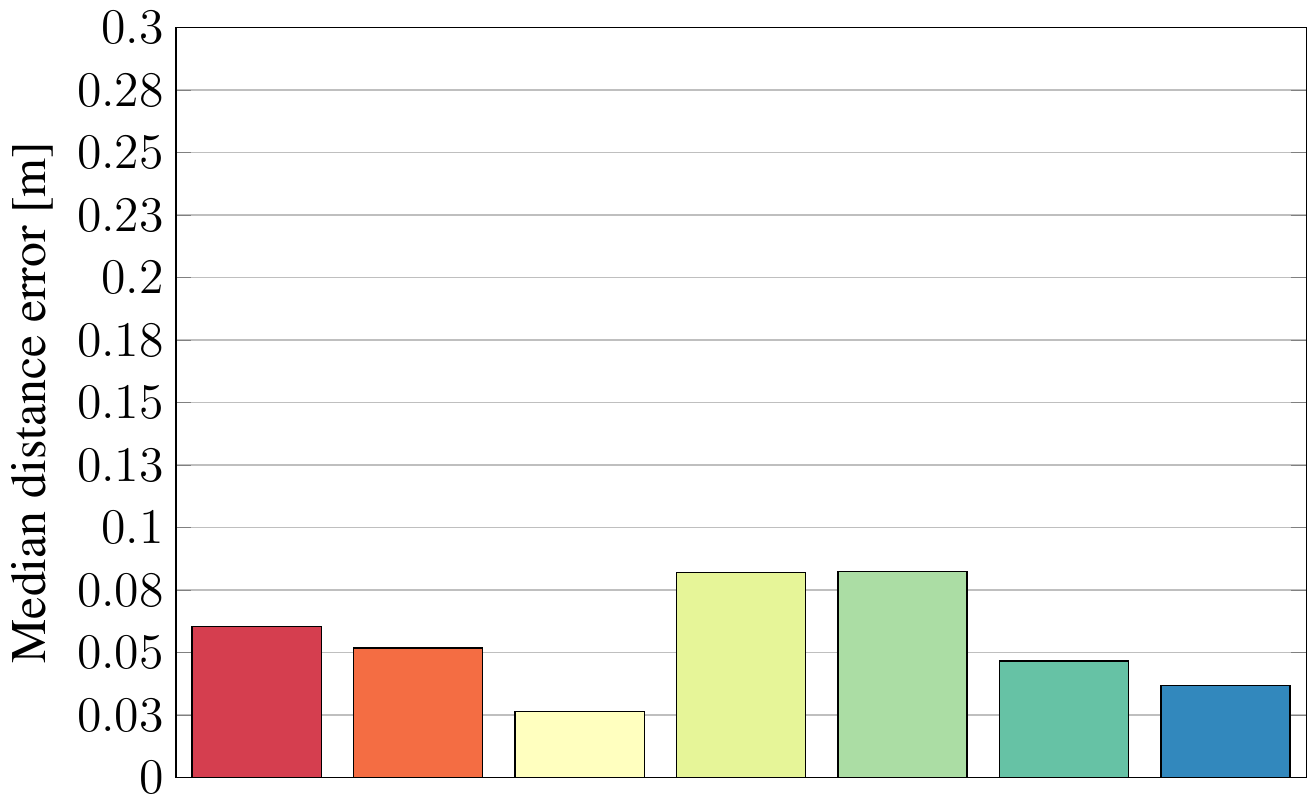}
		\hspace{\fighspace}
		\includegraphics[width=\figsizekf\textwidth]{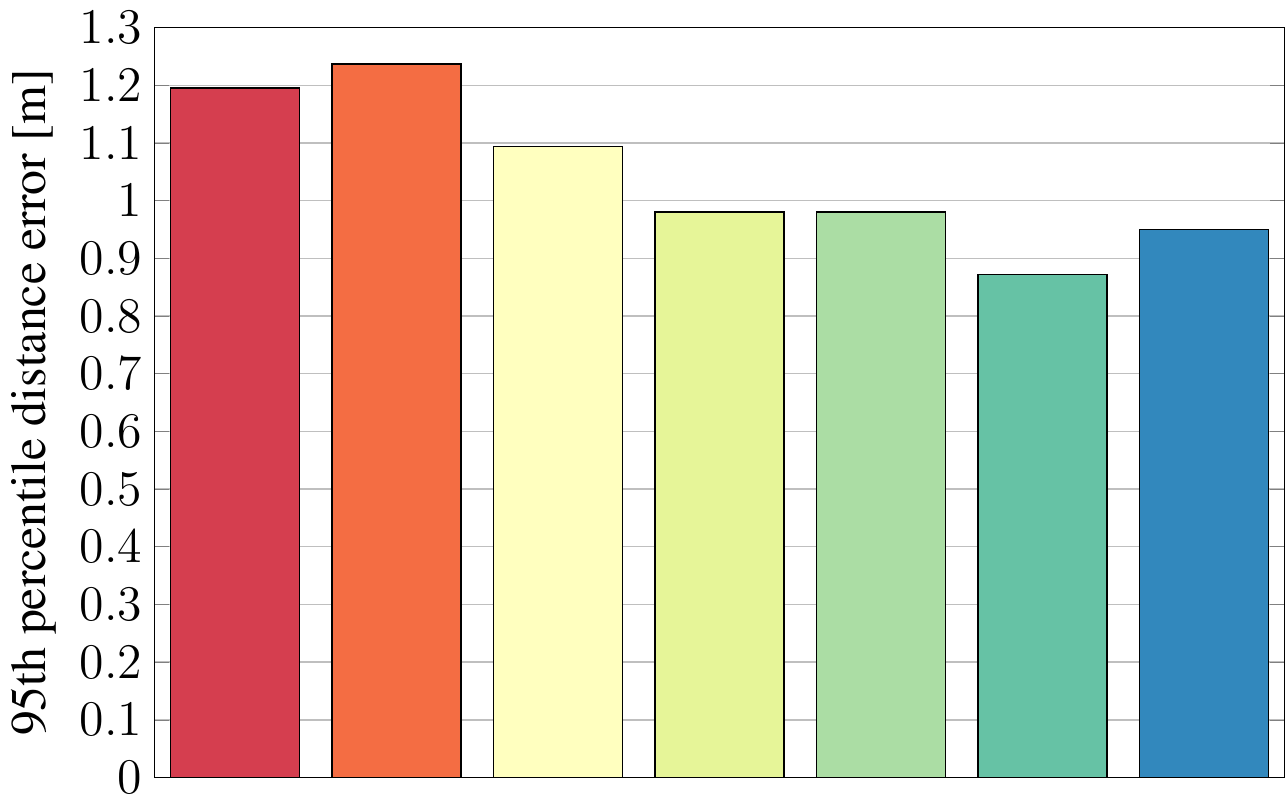}
	}
	\\
	\subfloat[]
	{
		\includegraphics[width=\figsizef\textwidth]{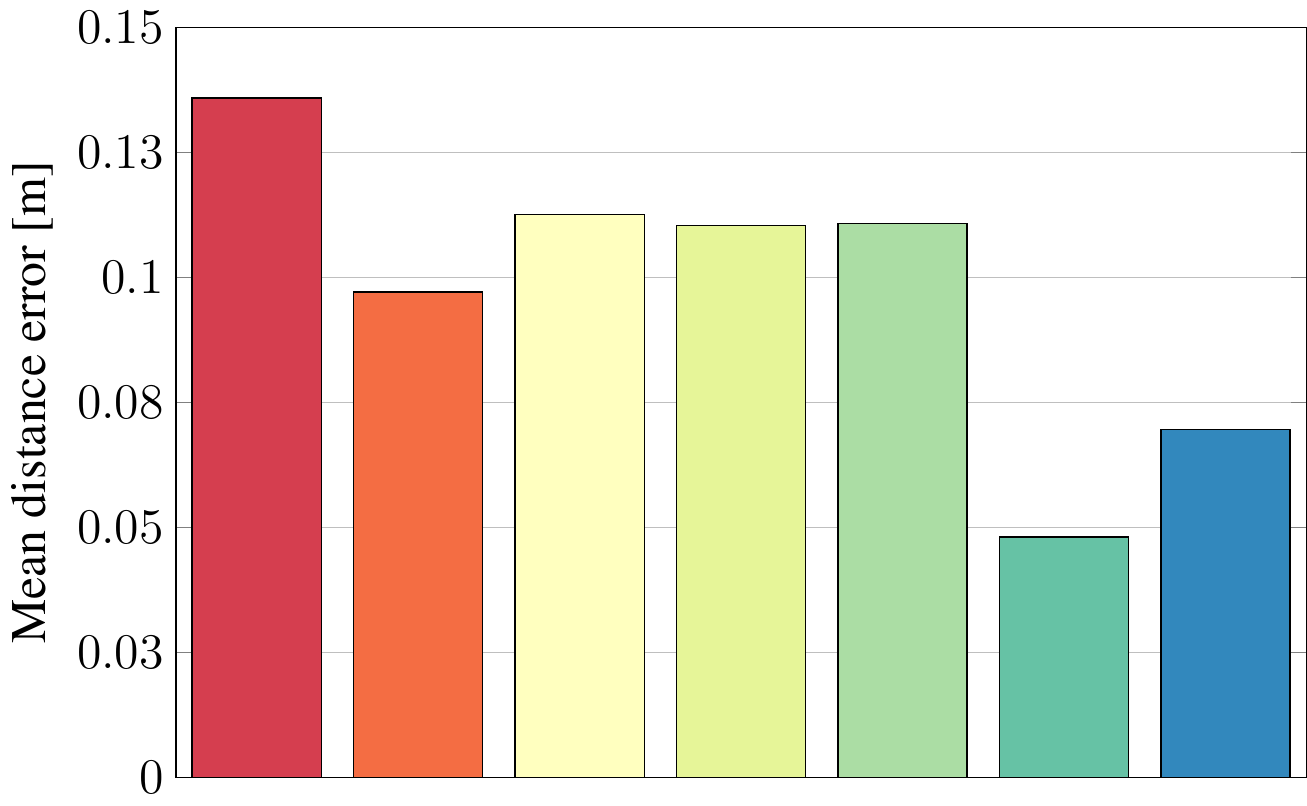}
		\hspace{\fighspace}
		\includegraphics[width=\figsizef\textwidth]{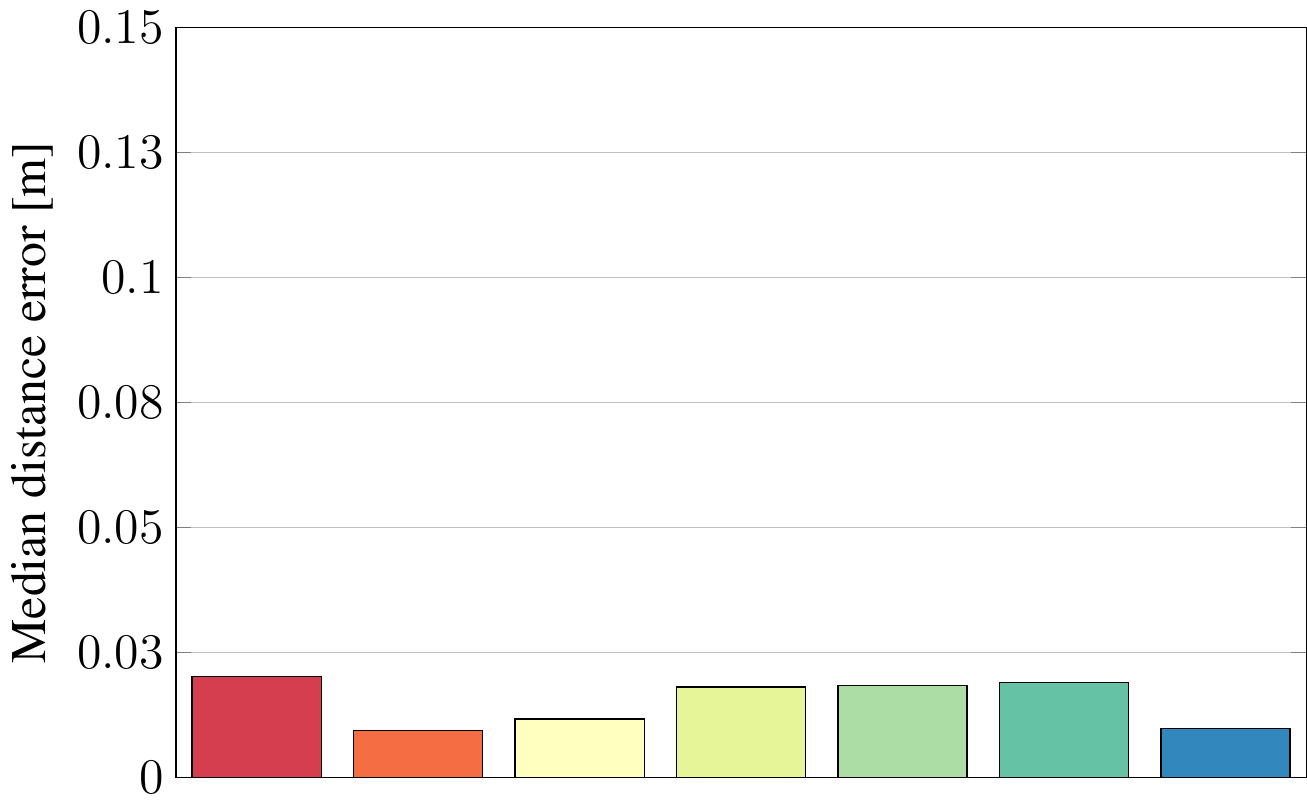}
		\hspace{\fighspace}
		\includegraphics[width=\figsizekf\textwidth]{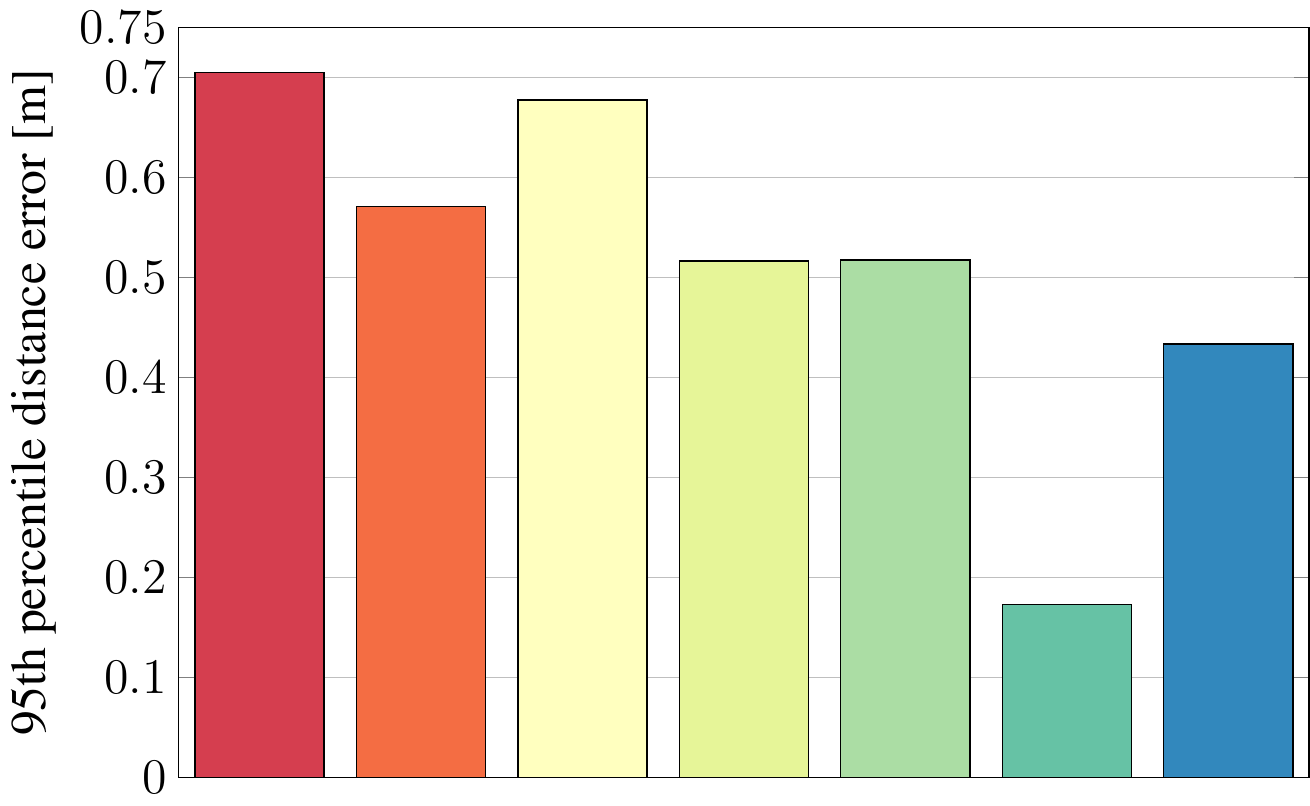}
	}
	\caption{Bar plots showing mean distance error (left), median distance error (middle), and the $95$th percentile distance error (right) evaluated on the test set for three multi-antenna scenarios: (a) single-point LoS lab scenario, (b) single-point non-LoS living room scenario, and (c) single-point non-LoS kitchen scenario.}
	\label{fig:barmultiantenna}	
\end{figure*}

\begin{figure*}[tp]
	\center
	\begin{minipage}{\textwidth}
		\centering
		~~\includegraphics[width=0.915\textwidth]{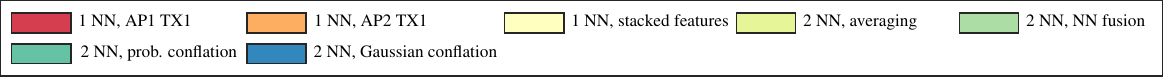}
	\end{minipage}\\[0.2cm]
	\subfloat
	{
		\includegraphics[width=\figsizef\textwidth]{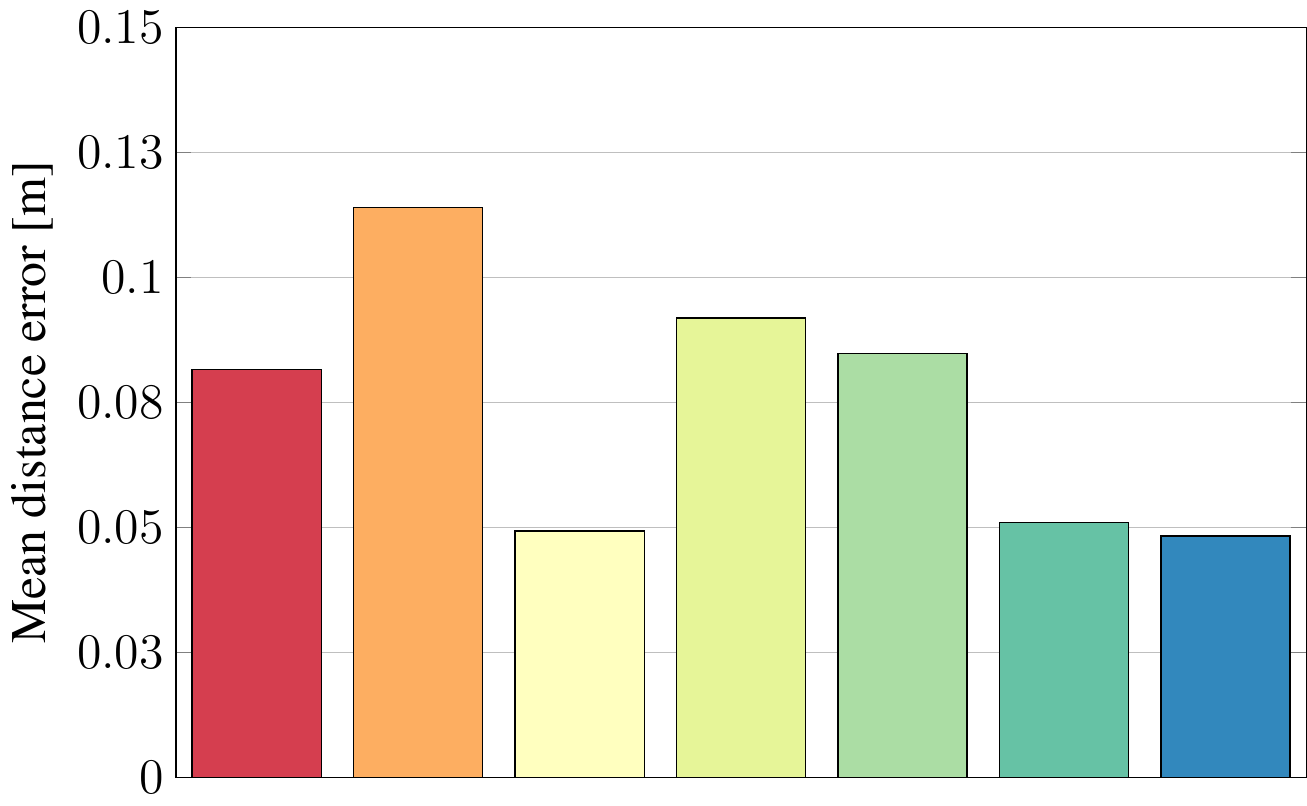}
		\hspace{\fighspace}
		\includegraphics[width=\figsizef\textwidth]{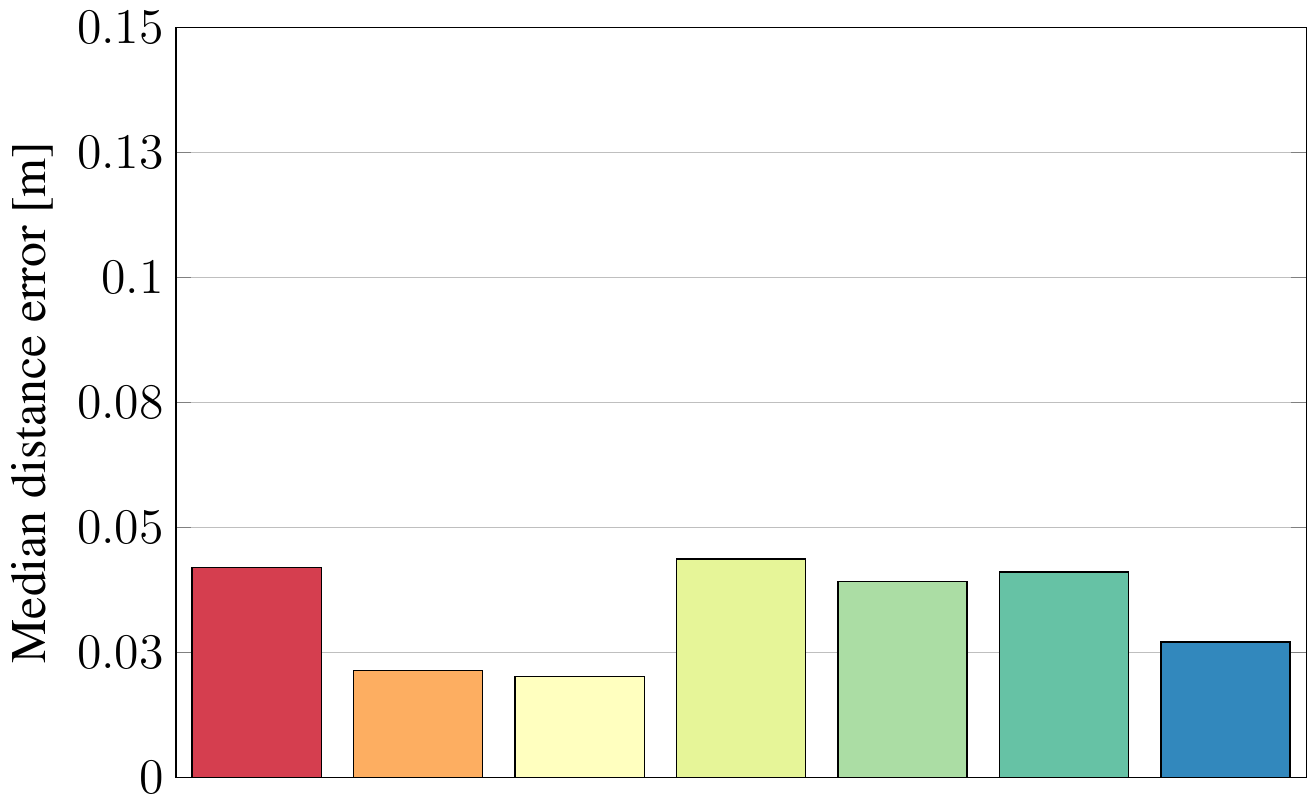}
		\hspace{\fighspace}
		\includegraphics[width=\figsizekf\textwidth]{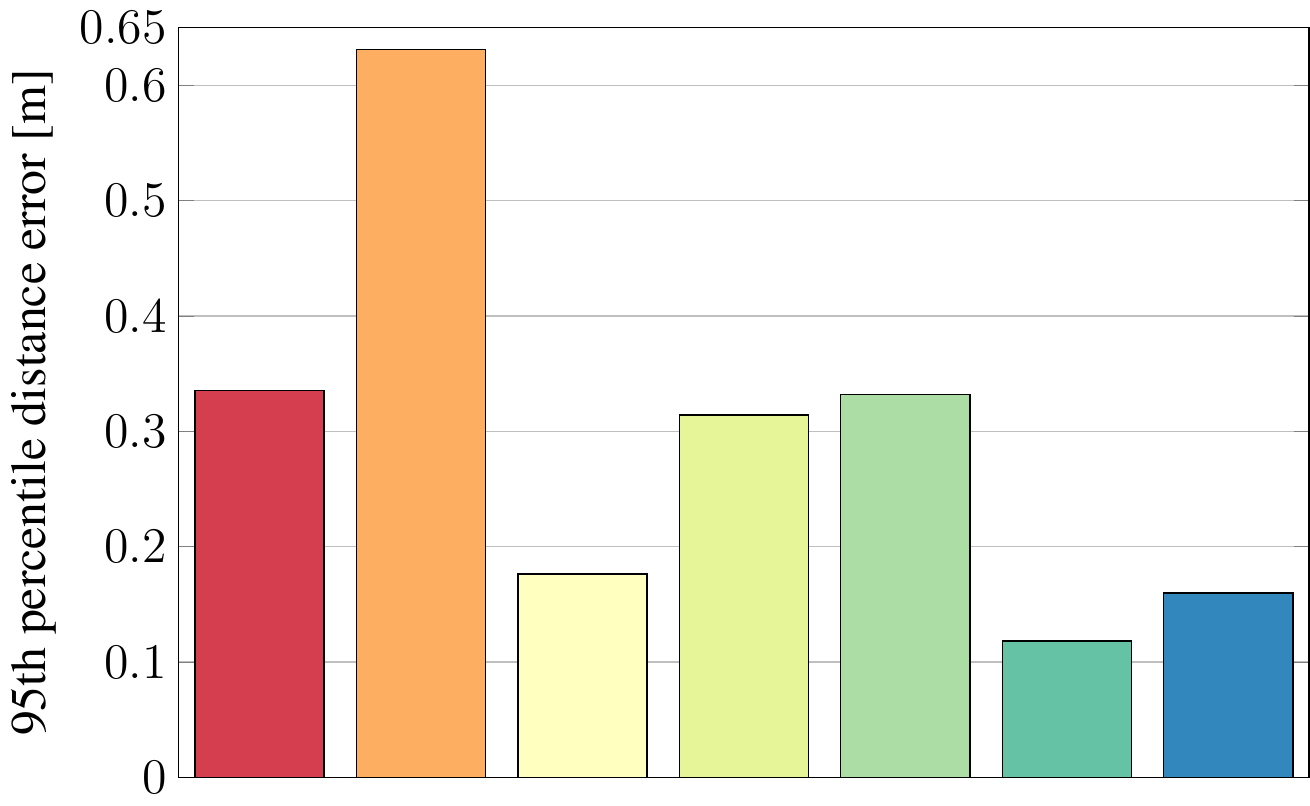}
	}
	\caption{Bar plots showing mean distance error (left), median distance error (middle), and the $95$th percentile distance error (right) evaluated on the test set for the single-antenna multi-point LoS living room scenario detailed in \fref{sec:measurementscenarios}.}
	\label{fig:barmultipoint}	
\end{figure*}

\begin{figure*}[tp]
	\center
	\begin{minipage}{\textwidth}
		\centering
		~~\includegraphics[width=0.915\textwidth]{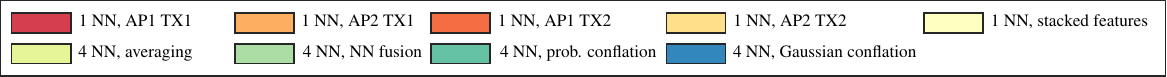}
	\end{minipage}\\[0.2cm]
	\subfloat
	{
		\includegraphics[width=\figsizef\textwidth]{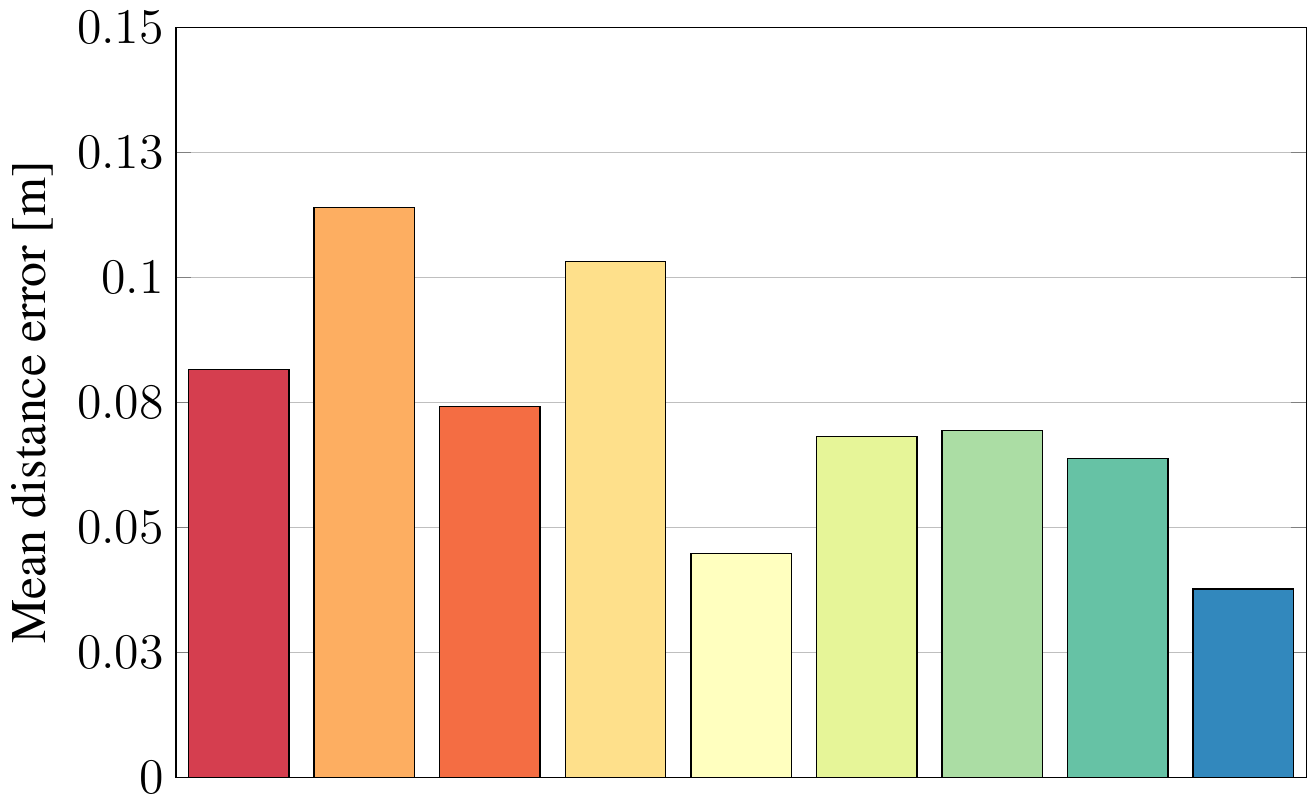}
		\hspace{\fighspace}
		\includegraphics[width=\figsizef\textwidth]{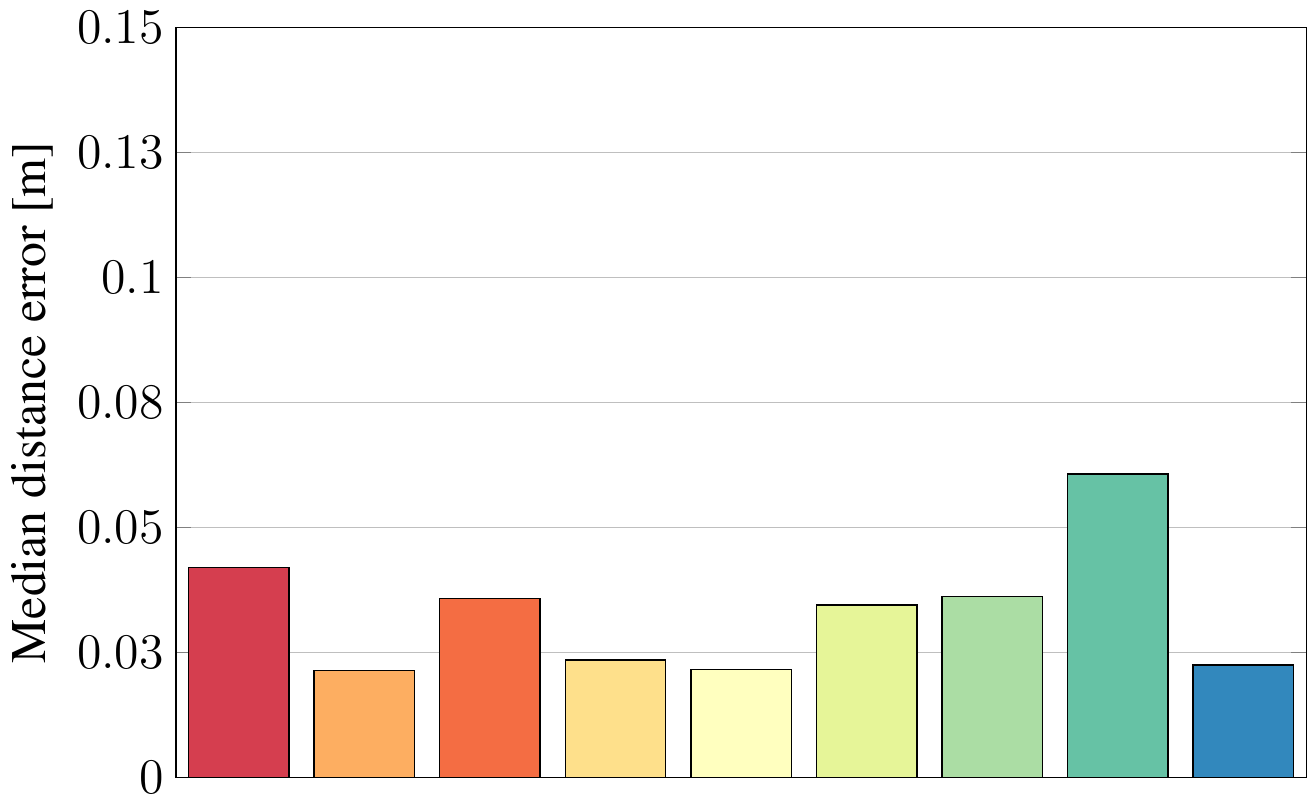}
		\hspace{\fighspace}
		\includegraphics[width=\figsizekf\textwidth]{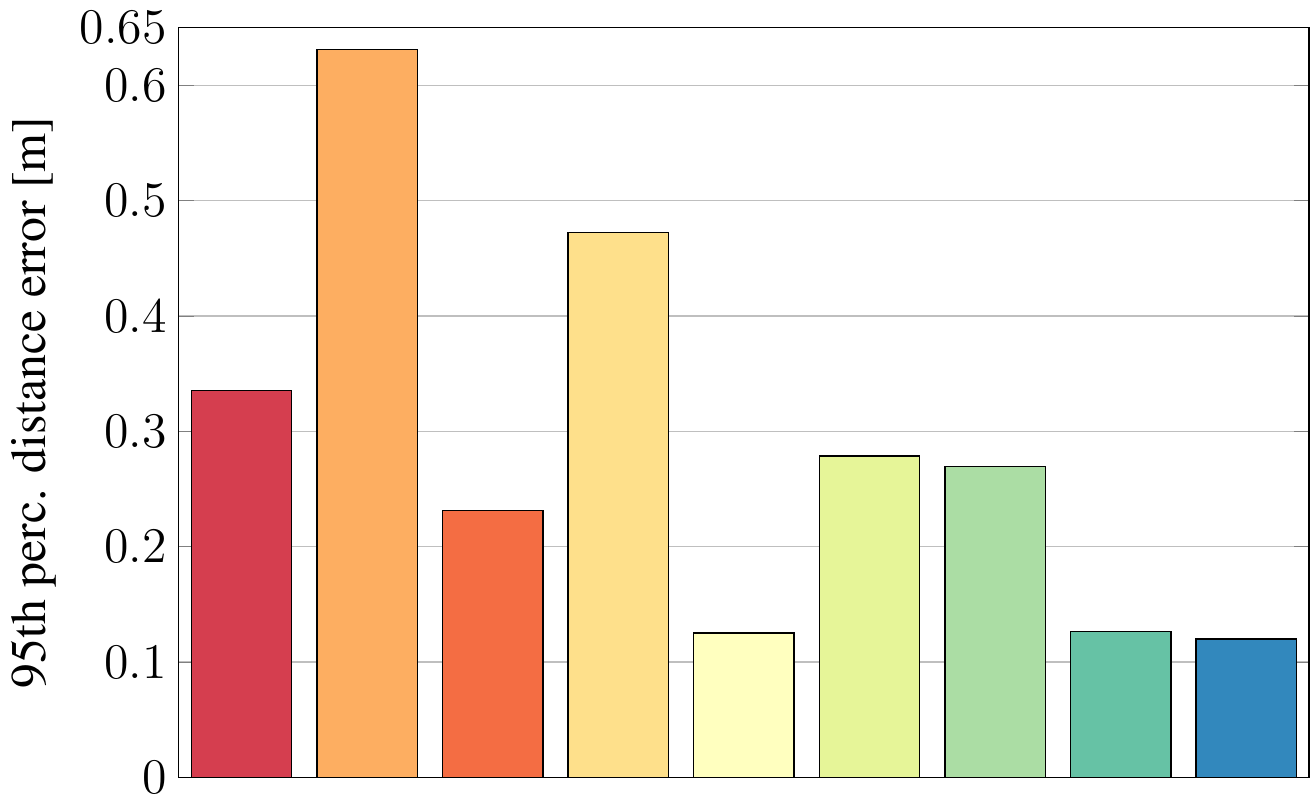}
	}
	\caption{Bar plots showing mean distance error (left), median distance error (middle), and the $95$th percentile distance error (right) evaluated on the test set for the multi-antenna multi-point LoS living room scenario detailed in \fref{sec:measurementscenarios}.}
	\label{fig:barmultipointmultiantenna}	
\end{figure*}

\newcommand{\figboxscale}{0.22}
\newcommand{\figboxhspace}{0.1}
\begin{figure*}[tp]
	\center
	\subfloat[]	
	{
		\includegraphics[width=\figboxscale\textwidth]{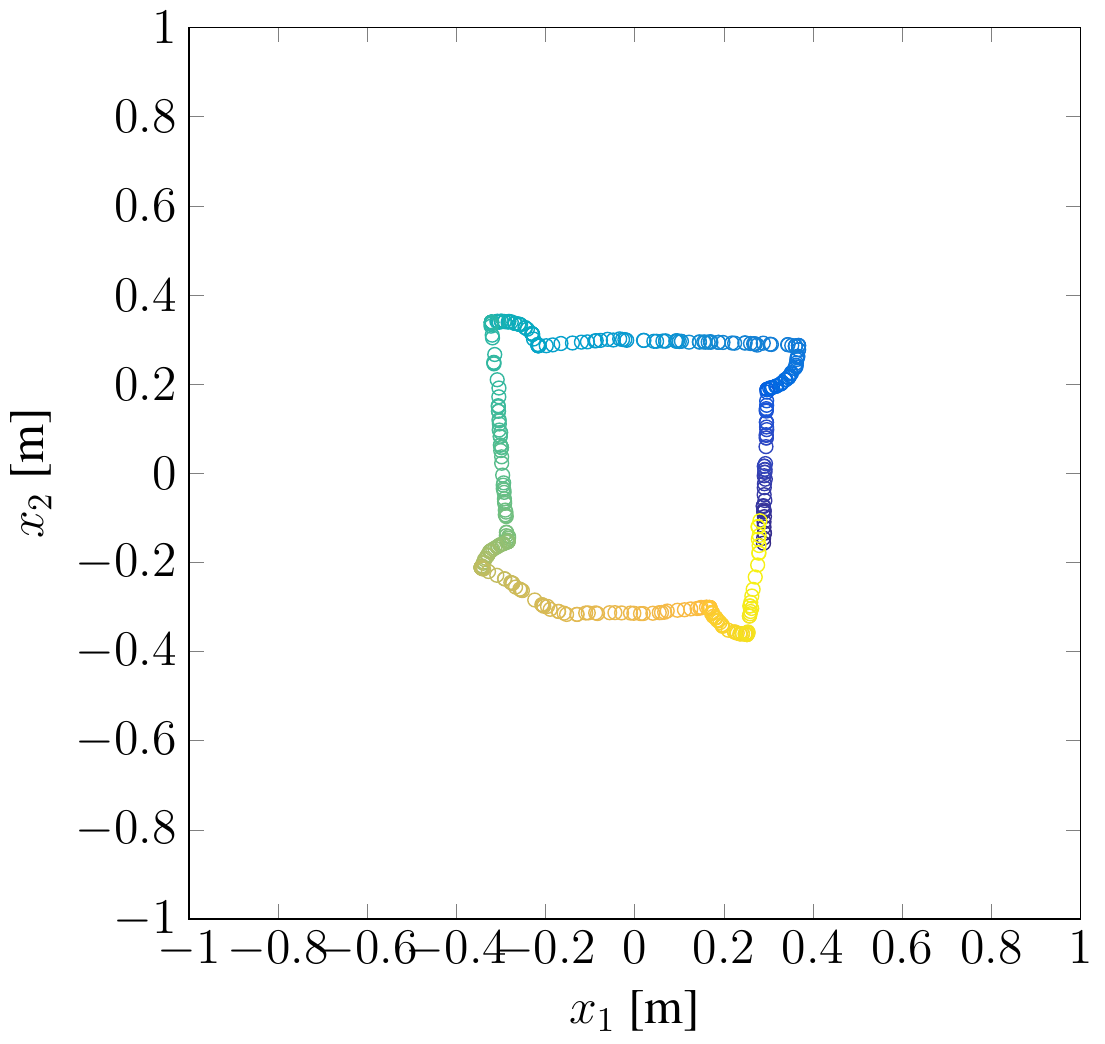}
	}
	\hspace{\figboxhspace cm}
	\subfloat[]
	{
		\includegraphics[width=\figboxscale\textwidth]{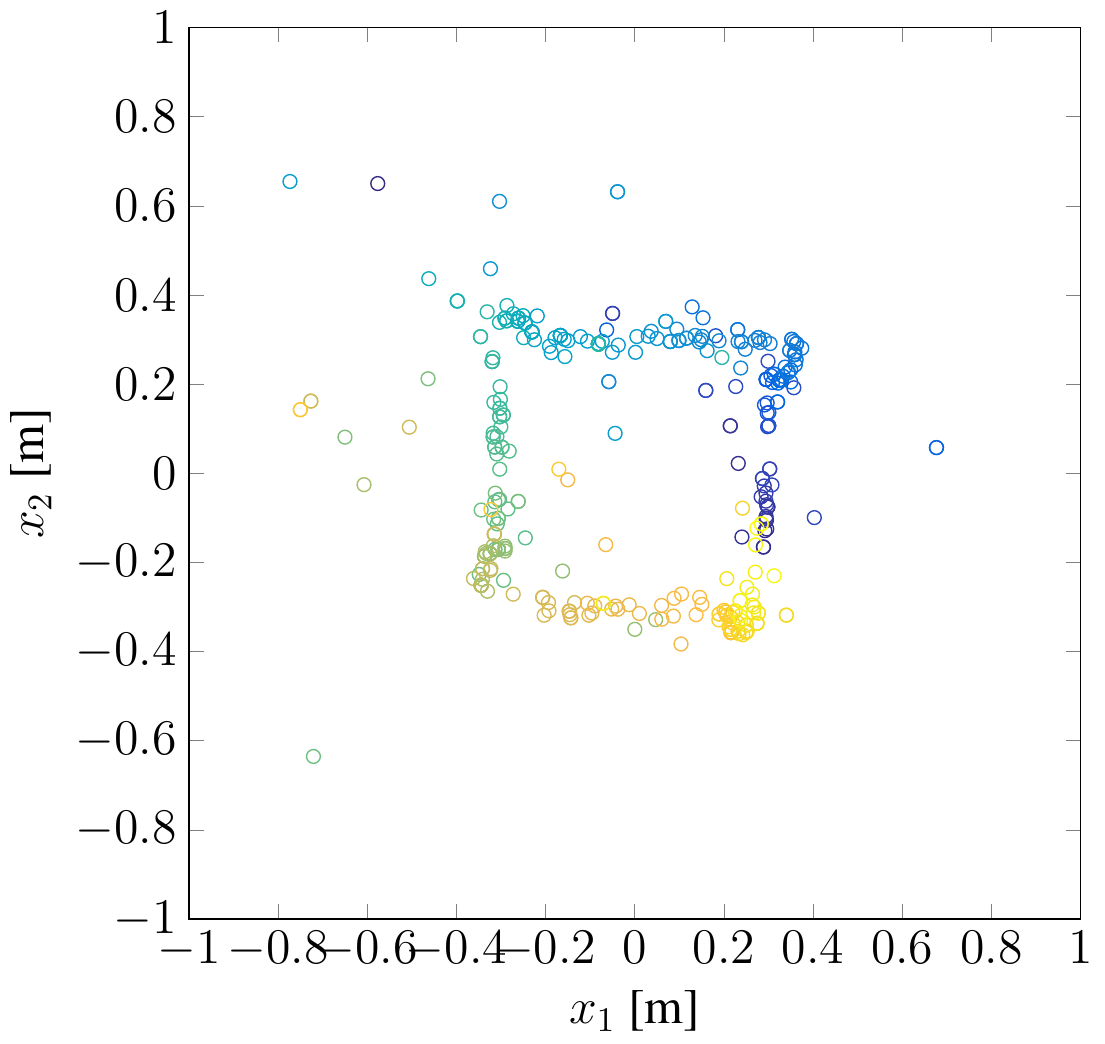}
	}
	\hspace{\figboxhspace cm}
	\subfloat[]	
	{
		\includegraphics[width=\figboxscale\textwidth]{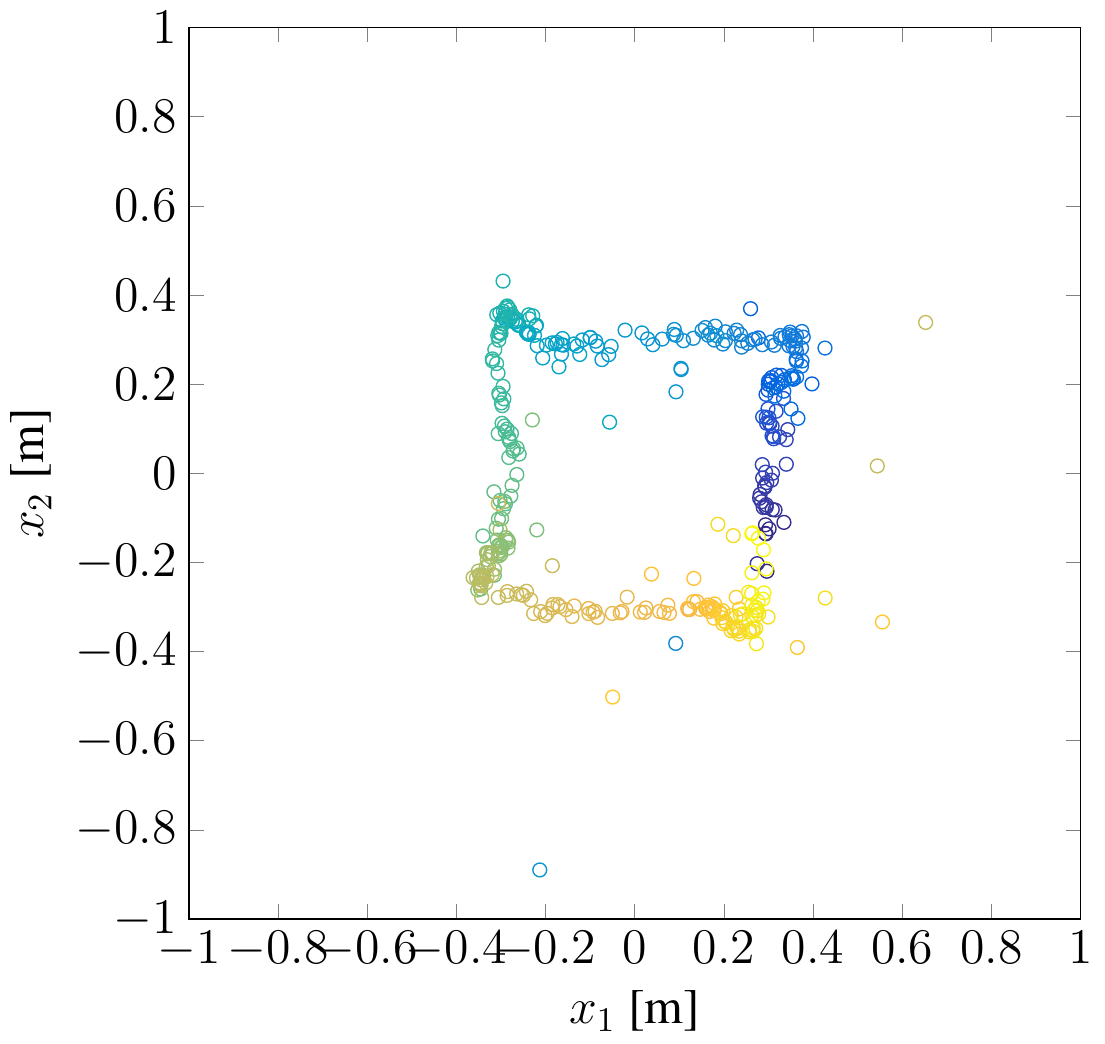}
	}
	\hspace{\figboxhspace cm}
	\subfloat[]	
	{
		\includegraphics[width=\figboxscale\textwidth]{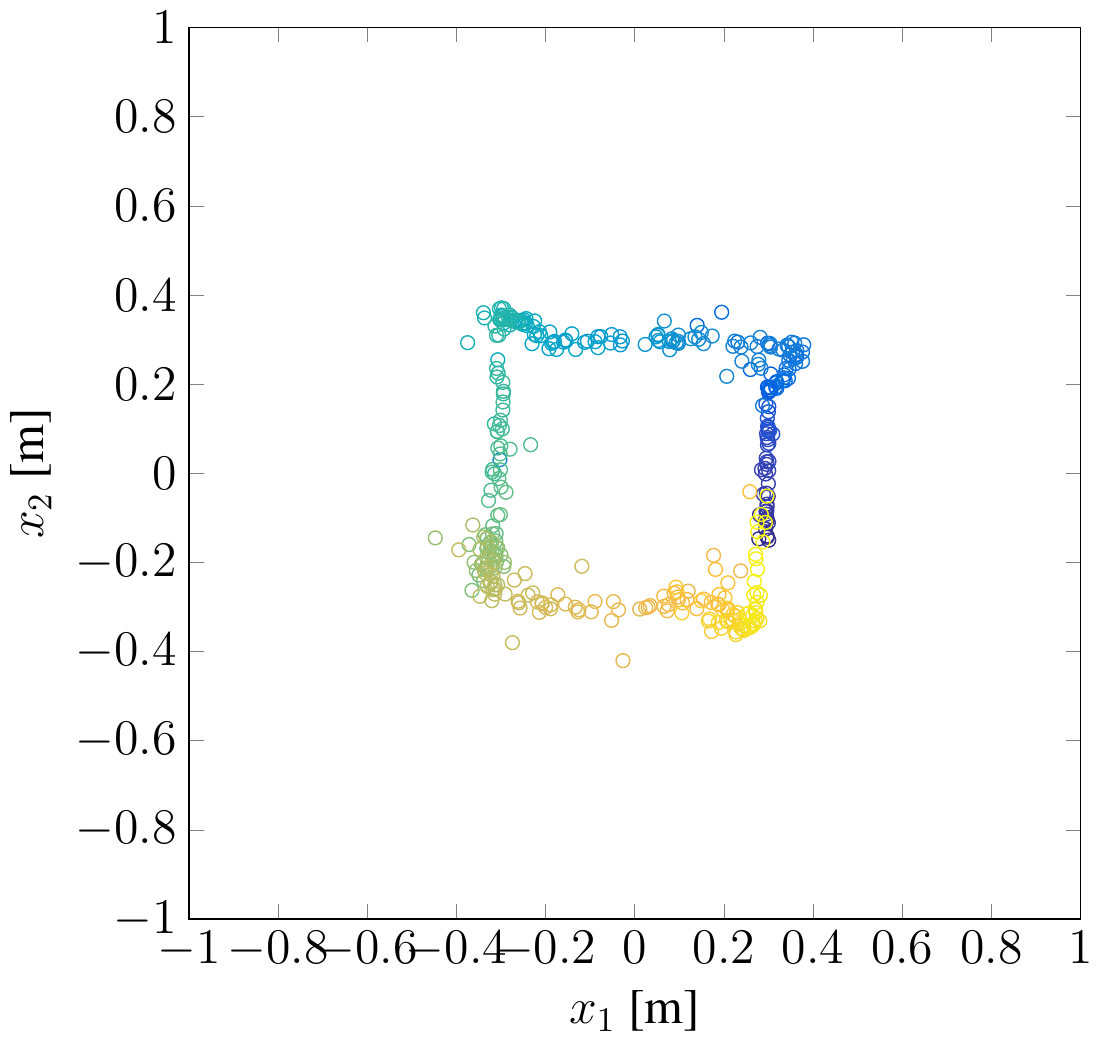}
	}
	\caption{Ground-truth locations (a) and estimated location (b), (c), and (d) for the multi-point LoS living room scenario detailed in \fref{sec:measurementscenarios}. (b) shows the performance of using a single NN from AP2 and TX1 ($\textit{MDE}=11$cm); (c) shows the performance of 1 NN with stacked features from 2 APs and 2 TX antennas ($\textit{MDE}=4.9$cm); and (d) shows the performance of 4 NNs with Gaussian conflation ($\textit{MDE}=3.7$cm).}
	\label{fig:multipointfusionresults}	
\end{figure*}

\section{Conclusions}
\label{sec:conclusions}
We have proposed CSI-based indoor positioning methods that are able to fuse one or multiple probability maps which describe the UEs' positions.
We have used a NN-based positioning pipeline that takes in features designed for MIMO-OFDM-based systems and are robust to typical hardware impairments, and generates probability maps.
We have proposed three different fusion methods for the computed probability maps, which reduce the amount of data to be transferred to a centralized processor that estimates UE position.
To demonstrate the effectiveness of the proposed positioning methods, we have evaluated our methods on four real-world indoor positioning datasets, which include multi transmit-antenna and multi AP scenarios.
Our comparison reveals three facts: (i) Indoor position accuracy of a few centimeters is possible from IEEE 802.11ac measurements, (ii) simple probability fusion techniques can significantly improve positioning accuracy while reducing the amount of data to be transported to a centralized processor, and (iii) multi-point probability fusion does not require accurate synchronization between the APs. 
We believe that the proposed probability fusion approach paves the way for other positioning systems or scenarios in which multiple features from different sensors are available.

There are many avenues for future work. First and foremost is the exploration of channel-charting based methods that (i) mitigate the need for dedicated CSI and ground-truth position measurement campaigns and (ii) generalize to the unseen positions.
Second is the development of accurate positioning pipelines that fuse multiple sensor modalities (besides CSI) which is part of ongoing work. 
Third is the evaluation of our approach in larger indoor spaces with multiple rooms and with a larger number of distributed APs.
Last is the acquisition of datasets with 3D position information and more (and controlled) dynamic changes in the environment. 

\section{Acknowledgments}
The authors would like to thank O. Casta\~neda and B.~Rappaport for discussions on CSI-based positioning using neural networks. We thank for D.\ Nonaca for useful comments on a preprint of this paper. We also thank Prof.\ K.~Petersen for allowing  us to use the Vicon positioning system \cite{vicon} and the lab space at Cornell University shown in \fref{fig:floorplans}(a).

\balance


\balance

\end{document}